\documentclass[12pt]{article}%%%%%%%%%%%%%%%%%%%%%%%%%%%%%%%%%%%%%%%%%%%%%%%%%%%%%%%%%%%%%%%%%%%%%%%%%%%%%%%%%%%%%%%%%%%%%%%%%%%%%%%%%%%%%%%%%%%%%%%%%%%%%%%%%%%%%%%%%%%%%%%%%%%%%%%%%%%%%%%%%%%%%%%%%%%%%%%%%%%%%%%%%%%%%%%%%%%%%%%%%%%%%%%%%%%%%%%%%%%%%%%%%%%%%%%%%%%%%%%%%%%%%%%%%%%%
\usepackage{hyperref}
\usepackage{amssymb}
\usepackage{amsmath}
\usepackage{graphicx}
\catcode`@=11
\renewcommand{\section}{\@startsection{section}{1}{0pt}{\medskipamount}
{\medskipamount}{\large\bf}}
\numberwithin{equation}{section}
\catcode`@=12

\def\beq{\begin{eqnarray}}
\def\eeq{\end{eqnarray}}
\def\ln{\,\mbox{ln}\,}

\def\im{\textrm{i}}

\def\sfrac#1#2{{\textstyle\frac{#1}{#2}}}
\def\={\ =\ }
\def\und{\qquad\textrm{and}\qquad}

\def\vp{\varepsilon}

\def\Ga{\Gamma}

\begin{document}

\begin{titlepage}
\setcounter{page}{0}

\vskip 2.0cm

\begin{center}

{\LARGE\bf
On Gauge Independence for Gauge Models with Soft Breaking of BRST Symmetry}

\vspace{18mm}

{\Large
Alexander Reshetnyak${}^{\dagger, \ast}$
}

\vspace{8mm}

\noindent ${}^\dagger${\em
Institute of Strength Physics and Material Science  \\
Siberian Branch of Russian Academy of Sciences,\\
 Akademicheskii av.\ 2/4, 634021 Tomsk, Russia}

\noindent ${}^\ast${\em
Tomsk State Pedagogical University,\\
Kievskaya St.\ 60, 634061 Tomsk, Russia}

\vspace{4mm}

\vspace{18mm}

\begin{abstract}
\noindent A consistent quantum treatment of general gauge theories
with an arbitrary gauge-fixing in the presence of soft breaking of
the BRST symmetry in the field-antifield formalism is developed.
It is based on a gauged (involving a field-dependent parameter)
version of finite BRST transformations. The prescription allows
one to restore the gauge-independence of the effective action at
its extremals and therefore also that of the conventional
$S$-matrix for a theory with BRST-breaking terms being additively
introduced into a BRST-invariant action in order to achieve a
consistency of the functional integral. We demonstrate the
applicability of this prescription within the approach of
functional renormalization group to the Yang--Mills and gravity
theories. The Gribov--Zwanziger action and the refined
Gribov--Zwanziger action for a many-parameter family of gauges,
including the Coulomb, axial and covariant gauges, are derived
perturbatively on the basis of finite gauged BRST transformations
starting from Landau gauge. It is proved that gauge theories with
soft breaking of BRST symmetry can be made consistent if the
transformed BRST-breaking terms satisfy the same soft BRST
symmetry breaking condition in the resulting gauge as the
untransformed ones in the initial gauge, and also without this
requirement.
\end{abstract}

\end{center}

%\vspace{12mm}
\vfill
\noindent{\sl Email:} \ reshet@ispms.tsc.ru
%% Covariant and sum-over-histories quantization
11.10.Hi, \ %% Renormalization group evolution of parameters
11.15.-q, \ %% Gauge field theories
11.15.Bt, \ %% General properties of perturbation theory
11.15.Tk  \ %% Other nonperturbative techniques
11.30.Pb

\end{titlepage}

%%%%%%%%%%%%%%%%%%%%%%%%%%%%%%%%%%%%%%%%%%%%%%%%%%%%%%%%%%%%%%

\section{Introduction}\label{intr}

\noindent

The contemporary progress in high-energy physics and quantum field
theory is strongly connected with the non-pertubative features of
quantum theories. The electroweak and strong interactions are
described by the Standard Model, in which Quantum Chromodynamics
(QCD) is a constituent, and there are no experimental facts in
conflict with QCD. While the Standard Model has been justified by
the discovery of the Higgs boson, the problem of consistency in
QCD is far from its solution, especially in view of the
confinement phenomenon. The Lagrangian of QCD (and generally that
of the Standard Model) belongs to the class of non-Abelian gauge
theories~\cite{books}, \cite{Bogolyubov}, \cite{FaddeevSlavnov}.
It is well known that the BRST symmetry~\cite{brst}, being a
special global fermionic descendant of gauge invariance, plays a
fundamental role in quantum field theory, since the fundamental
interactions of Nature, together with gravity and perhaps some yet
unknown forces, can be described in terms of gauge theories. The
covariant quantization of Yang--Mills theories by means of the
Faddeev--Popov procedure cannot be realized correctly, even
perturbatively, for the entire spectrum of the momenta
distribution due to the well-known Gribov problem \cite{Gribov} in
the deep infra-red region for gauge fields, once a gauge condition
has been imposed differentially \cite{Singer}, since there remains
an infinitely large number of discrete gauge copies even after
gauge-fixing.\footnote{There are some other recently suggested
methods of solving the Gribov problem in a consistent way: first,
the procedure of imposing an algebraic (instead of a differential
one) gauge on auxiliary scalar fields in a theory which is
non-perturbatively equivalent to the Yang-Mills theory with the
same gauge group \cite{Slavnoveq}; second, the procedure of
averaging over the Gribov copies with a non-uniform weight in the
functional integral and the replica trick \cite{Serreau}.}

In order to fix the residual gauge freedom, Gribov has undertaken
a detailed study of the Coulomb gauge and suggested a restriction
of the domain of functional integration for gauge fields to the
so-called first Gribov region, which has been effectively
incorporated into the functional measure as the Heaviside
$\Theta$-function, thus realizing the ``no-pole'' condition for
the ghost propagator. Effectively, this restriction can be
implemented, in the Landau gauge with a hermitian Faddeev--Popov
operator, by a special addition introduced to the standard
Faddeev--Popov action and known as the Gribov--Zwanziger
functional \cite{Zwanziger1}, \cite{Zwanziger2}. However, this
addition is not gauge-invariant and is therefore non-invariant
under the original BRST transformations.

The idea of using the Zwanziger action in order to take account of
gauge field configurations has introduced to the path integral of
the Yang--Mills theory the entire spectrum of frequencies, which
has been examined in a number of papers~\cite{Sorellas} based on
the breakdown of BRST symmetry in Yang--Mills theories. Notice
that until now the considerations \cite{Zwanziger1},
\cite{Zwanziger2}, \cite{Sorellas} of the Gribov horizon in
Yang--Mills theories have been carried out basically in the Landau
gauge. The analytical proof \cite{Gribov} of the presence of
Gribov copies in the physical spectrum has been confirmed by
lattice simulations in some QCD models, such as the $SU(2)$
gluodynamics (see, e.g., \cite{lattice}, \cite{lattice2} and
references therein), which is an expected result due to the
discovery of field configurations within the same Landau gauge
condition for the Faddeev--Popov action adopted to lattice
calculations. For the sake of completeness, notice the study
\cite{SS} of the Gribov problem (beyond the Landau gauge) in
covariant $R_\xi$ gauges for a small value of the gauge parameter
$\xi$ for an approximation of the quantum action being quadratic
in the fields; let us also notice the proposal of a new form of
the horizon functional in $R_\xi$ gauges \cite{rl} in the maximal
Abelian gauges \cite{MAG}, \cite{MAG1}, in the Coulomb gauge
\cite{HFZwanziger}, and on a curved Riemannian background to study
the influence of the curvature tensor on changing the size of the
Gribov region \cite{curvedGribov}.

There is a large freedom in the choice of admissible gauges used
to obtain a correct path integral in Yang--Mills theories with
account of the Gribov problem; it is also well known that the
Green functions are gauge-dependent; however, this dependence has
such a specific character that it should be cancelled in physical
combinations such as the S-matrix. Contemporary proofs of the
gauge-independence of the S-matrix in Yang--Mills theories are
based on the BRST symmetry, see, e.g., \cite{LT}, and also apply
to more general gauge theories \cite{VLT}. There arises an
immediate problem of consistency for a gauge theory in case the
BRST invariance of the resulting quantum action (such as the
Gribov--Zwanziger action) turns out to be broken. The study of
this problem has been initiated by \cite{llr}, \cite{lrr}. These
studies investigated both Yang--Mills and general gauge theories,
such as supergravity, superstrings with open algebras, and
higher-spin fields as reducible gauge theories, see, e.g.,
\cite{hspin1}, with an introduction of so-called soft breaking of
the BRST symmetry (under the natural assumption of the existence
of the Gribov horizon and Gribov--Zwanziger functional for any
theory with a non-Abelian gauge algebra) and achieved their
results on a basis of the field-antifield method \cite{BV1},
\cite{BV2}. Namely, in \cite{llr}, \cite{lrr}, it was shown (with
some peculiar features studied in \cite{rr}) that the
gauge-independence of the effective action for a gauge theory with
soft BRST symmetry breaking on the mass shell requires the
fulfillment of a quite strong condition for the BRST symmetry
breaking term, and therefore we  come to the conclusion
\cite{llr}: ``It is argued that gauge
theories with a soft breaking of BRST symmetry are inconsistent.''
The same statement has been shown to take place in
the Gribov--Zwanziger theory with the $R_\xi$-gauge.

As a next step in solving the problem of determining the horizon
functional for the Yang--Mills theory in gauges beyond the Landau
gauge, there is the recent concept of so-called finite
field-dependent BRST transformations \cite{ll1}, earlier used in
the infinitesimal form \cite{BV1}, \cite{BV2} in order to
establish the gauge-independence of the vacuum functional, but now
explicitly constructed to relate the Faddeev--Popov action in the
Landau gauge and in the covariant $R_\xi$-gauge.
 The concept of field-dependent BRST transformations, first
suggested \cite{JM} in a finite (however, different) form (see
also \cite{JM1}, \cite{Upadhyay1}) permits one to obtain
perturbatively an explicit form of the Gribov horizon functional
in the $R_\xi$-gauge \cite{ll2}, starting from its form in the
Landau gauge. This provides a different perspective of the problem
of gauge-dependence for the Gribov--Zwanziger theory and allows
one to revisit this problem more generally in a gauge theory with
soft BRST breaking symmetry.

In this work, we present a study of gauge-dependence in general
gauge theories with soft breaking of the BRST breaking and
develop, for this purpose, a concept of gauged (equivalently,
field-antifield-dependent) BRST transformations. We present an
explicit calculation of the Jacobian of the corresponding change
of field-antifield variables in the partition function, to
determine and solve a non-linear equation for an unknown
field-antifield-dependent odd-valued parameter $\Lambda$. We
establish a coincidence of the vacuum functional without a BRST
broken term in a gauge determined by a gauge Fermion $\Psi$ with
the vacuum functional in a different gauge determined by
$\Psi+\overline{\Delta} \Psi$. On this basis, we examine the
properties of the average effective action within the approach of
the functional renormalization group to the Yang--Mills and
gravity theories. We also suggest the Gribov--Zwanziger horizon
functional for a many-parameter family of linear gauges, including
the Coulomb, the axial, and the $R_\xi$ gauges, used in
non-Hermitian Faddeev--Popov operators.

The paper is organized as follows. In Section~\ref{gBRSTtrans}, we
introduce the concept of finite gauged (field-dependent) BRST
symmetry transformations and investigate the related change of
variables in the functional integral for general gauge theories in
the field-antifield formalism. In Section~\ref{gengd}, we use the
field-dependent BRST transformations to formulate the study of
gauge-dependence for the generating functionals of Green's
functions for a general gauge theory with BRST-broken terms in
arbitrary gauges (using a suitable regularization scheme); we also
formulate the main result of this study. An application of the
general results to the functional renormalization group approach
to the Yang--Mills and gravity theories is considered in
Section~\ref{FRGandPV}. In Section~\ref{rGZ}, we examine different
choices for gauged BRST transformations in order to find the form
of the Gribov--Zwanziger action and of the refined
Gribov--Zwanziger action in a many-parameter family of gauges
including the Coulomb, axial, Landau and covariant gauges,
starting from the Landau gauge. Finally, in Section~\ref{concl} we
discuss some issues and perspectives related to the suggested
procedure. In Appendix~\ref{AppA}, we analyze the existence of a
solution to a non-linear functional equation for an unknown
field-antifield-dependent odd-valued parameter $\Lambda$, which
establishes the coincidence of the vacuum functionals in different
gauges.

We use the condensed notation of DeWitt~\cite{DeWitt} and our
previous notation~\cite{llr}, \cite{lrr}, \cite{rl}. Derivatives
with respect to sources and antifields are taken from the left,
while those with respect to fields are taken from the right. Left
derivatives with respect to fields are labelled
by the subscript~``$l$''. The Grassmann parity of a quantity $A$ is denoted by $%
\varepsilon (A)$.

\section{Gauged BRST Symmetry Transformations}

\label{gBRSTtrans}

In this section, we recall the basic notions and properties of the
field-antifield formalism for general gauge theories. We also
introduce gauged (field-dependent) BRST transformations and
calculate the Jacobian for the change of variables determined by
these transformations.

\subsection{Overview of Field-antifield Formalism}

\label{generals} \noindent

As the initial point of our study, we consider a theory of gauge fields, $%
A^{i}$, $i=1,2,\ldots ,n$, with $\varepsilon (A^{i})=\varepsilon
_{i}$, determined by a classical action, $S_{0}=S_{0}(A)$,
invariant under infinitesimal gauge transformations $\delta
A^{i}=R_{\alpha _{0}}^{i}(A)\xi ^{\alpha _{0}}$ {for} $\alpha
_{0}=1,2,\ldots ,m_{0}$, implying the Noether identities
\begin{eqnarray}
\label{GIClassA}
S_{0,i}(A) R^i_{\alpha_0}(A)=0 ,\qquad 0<m_0<n .
\end{eqnarray}
Here, the gauge transformations are parameterized by $m_{0}$
arbitrary (usually supposed to be small) functions, $\xi ^{\alpha
_{0}}$, of the space-time coordinates, with $\varepsilon (\xi
^{\alpha _{0}})=\varepsilon
_{\alpha _{0}}$, whereas $S_{0,i}\equiv \delta S_{0}/\delta A^{i}$, while $%
R_{\alpha _{0}}^{i}(A)$ are the generators of the gauge transformations,
with $\varepsilon (R_{\alpha _{0}}^{i})=\varepsilon _{i}{+}\varepsilon
_{\alpha _{0}}$.

The generators may be dependent in the case $\mathrm{rank}\Vert
R_{\alpha _{0}}^{i}\Vert _{\mathcal{S}_{0,i}=0}=m<m_{0}$, implying
the presence of zero eigenvectors, $Z_{\alpha _{1}}^{\alpha
_{0}}(A)$, $\alpha
_{1}=1,...,m_{1}$, for the generators on the mass shell $\mathcal{S}%
_{0,i}=0 $, thus determining a reducible gauge theory, so that in the case $%
\mathrm{rank}\Vert Z_{\alpha _{1}}^{\alpha _{0}}\Vert _{\mathcal{S}%
_{0,i}=0}<m_{1}$ the eigenvectors should be dependent as well. Thus, an $L$%
-th-stage reducible gauge theory of the fields $A^{i}$ is determined by the
relations
 \begin{eqnarray}
\label{redth}
 && Z^{\alpha_{s-1}}_{\alpha_{s}}(A)Z^{\alpha_s}_{\alpha_{s+1}}(A)=  \mathcal{S}_{0,i}(A) K^{i\alpha_{s-1}}_{\alpha_{s+1}}(A),\texttt{ for } \alpha_{s+1}=1,...,m_{s+1}, s = 0,...,L-1,\\
 && \texttt{ and }\mathrm{rank}\|Z^{\alpha_{s-1}}_{\alpha_{s}}\|_{\mathcal{S}_{0,i}=0}< m_s,\quad  \mathrm{rank}\|Z^{\alpha_{L-1}}_{\alpha_{L}}\|_{\mathcal{S}_{0,i}=0}= m_L,\\
 \label{redth1}
 && \texttt{ where }Z^{\alpha_{-1}}_{\alpha_{0}}\equiv R^i_{\alpha_0}, \quad \vp(Z^{\alpha_s}_{\alpha_{s+1}})= \vp_{\alpha_s}+\vp_{\alpha_{s+1}}, \quad \vp(K^{i\alpha_{s-1}}_{\alpha_{s+1}})= \vp_i+\vp_{\alpha_{s-1}}+\vp_{\alpha_{s+1}}.
\end{eqnarray}
The total configuration space $\mathcal{M}$ of all the fields
$\{\Phi ^{A}\}$ in the BV method depends on the irreducible
\cite{BV1} or reducible \cite{BV2} nature of a given classical
gauge theory. In the case of an $L$-th stage reducible theory,
$\mathcal{M}$ is parameterized by the fields
\begin{equation}\label{field}
\Phi \ \equiv \ \{\Phi ^{A}\} = \{A^{i},C^{\alpha _{s}},{C}{}_{s^{\prime
}}^{\alpha _{s}},{B}{}_{s^{\prime }}^{\alpha _{s}}\},\quad s=0,...,L,\ s^{\prime
}=0,...,s,
\end{equation}%
with $\varepsilon (C^{\alpha _{s}},{C}{}_{s^{\prime }}^{\alpha _{s}},{B}%
{}_{s^{\prime }}^{\alpha _{s}})=$ $(\varepsilon _{\alpha
_{s}}+s+1,\varepsilon _{\alpha _{s}}+s+1,\varepsilon _{\alpha _{s}}+s)$, $%
\varepsilon (\Phi ^{A})=\varepsilon _{A}$ and the following ghost
number distribution:
\begin{equation*}
gh\big(A^i, C^{\alpha _{s}},\ {C}{}_{s^{\prime }}^{\alpha _{s}},\ {B}%
{}_{s^{\prime }}^{\alpha _{s}}\big)\ =\ \big(0,\ s+1,\ 2s'-s-1,\ 2s'-s\big),
\end{equation*}
which obeys an additive composition law when calculated on
monomials. Here, the respective classical, minimal-ghost,
antighost, extra-ghost and Nakanishi--Lautrup fields are
explicitly indicated in the BV method. For $L=0$, the gauge theory
is
irreducible, with $C^{\alpha _{0}},{C}{}_{0}^{\alpha _{0}}\equiv \overline{C%
}{}^{\alpha _{0}},{B}{}_{0}^{\alpha _{0}}\equiv {B}{}^{\alpha _{0}}$ being
the ghost, antighost and Nakanishi--Lautrup fields.

The BV method demands the introduction of an odd cotangent bundle
$\Pi T^{\ast }\mathcal{M}\equiv T^{\ast (0,1)}\mathcal{M}$,
usually known as the field-antifield space (for a more involved
geometry, based on the field-antifield formalism, see, e.g., Refs.
\cite{hudaners}, \cite{bt1},
 \cite{ashwarz},  \cite{lmr}, \cite{gmr},
\cite{reshrhg}, \cite{bb1}), where each field $\Phi ^{A}$ in
$\mathcal{M}$ has a corresponding antifield~$\Phi ^{\ast }\ \equiv
\Phi _{A}^{\ast }$,
\begin{equation}\label{afield}
\{\Phi _{A}^{\ast }\}\ =\ \{A_{i}^{\ast },C_{\alpha _{s}}^{\ast },{C}%
{}_{s^{\prime }{}\alpha _{s}}^{\ast },{B}{}_{s^{\prime }{}\alpha _{s}}^{\ast
}\},\,\,\,\mathrm{with}\,\,\,(\varepsilon ,gh)(\Phi _{A}^{\ast })=(\varepsilon _{A}{+}%
1,-1-gh(\Phi ^{A})).
\end{equation}

%%%%%%%%%%%%%%%%%%%%%%%%%%%%%%%%%%%%%%
In the total field-antifield space $\{\Phi ^{A},\Phi _{A}^{\ast
}\}$, one defines a bosonic functional, ${\bar{S}}={\bar{S}}(\Phi
,\Phi ^{\ast })$,
being a special extension of the classical action to $\Pi T^{\ast }\mathcal{M%
}$ with the boundary condition of vanishing antifields $\Phi
_{A}^{\ast }$ and Planck constant, ${\bar{S}}(\Phi ,0)\big|_{\hbar
=0}=\mathcal{S}_{0}$, with $gh(S)=0$, encoding the gauge algebra
functions and satisfying a quantum master equation (within the
class of gauge-invariant regularizations,
with $\Delta \bar{S}\sim \delta (0)\neq 0$ for a local $%
\bar{S}$) in two equivalent forms:%
\beq \label{MEBV}
{\Delta}\exp\left\{\frac{\im}{\hbar}{\bar S}\right\} =0 \Longleftrightarrow
\sfrac {1}{2} ({\bar S},{\bar
S})\=\im\hbar\,{\Delta}{\bar S}. \eeq
These equations are written in terms of a natural (in $\Pi T^{\ast }\mathcal{M}$)
odd Poisson bracket, $(\bullet ,\bullet )$, (known as the antibracket) and a nilpotent odd Laplacian, $%
\Delta $,
\begin{equation}\label{abr}
(\bullet ,\bullet )\ =\ \frac{\delta \bullet }{\delta \Phi ^{A}}\,\frac{%
\delta \bullet }{\delta \Phi _{A}^{\ast }}\ -\ \frac{\delta _{\mathit{r}}\bullet }{\delta
\Phi _{A}^{\ast }}\,\frac{\delta _{\mathit{l}
}\bullet }{\delta \Phi ^{A}},\qquad \Delta \ =\ (-1)^{\varepsilon _{A}}\frac{\delta _{%
\mathit{l}}}{\delta \Phi ^{A}}\;\frac{\delta }{\delta \Phi _{A}^{\ast }}.
\end{equation}%
We assume that formal manipulations with  $\Delta $ are
supported by a suitable regularization scheme. This is a nontrivial
requirement, since  $\Delta $ is not well-defined on local
functionals,\footnote{%
There is another proposal \cite{Kiselev} to define an odd
Laplacian, thus solving the problem of $\delta (0)$.} because for any
local functional $F$ one finds that $\Delta F\sim \delta (0)$. The
standard way to solve this problem is to
use a regularization similar to the dimensional one~\cite{Leib}, when $%
\delta (0)=0$. In this paper, just as in \cite{lrr}, we consider a more
general class of regularizations.

The quantum action is constructed as a special representative from the set
of solutions to the master equation (\ref{MEBV}) and is described by the
transformation
\beq \label{qaBV}
\exp\left\{\frac{\im}{\hbar}{ S_X}\right\} = \exp\left\{-[\Delta, X]\right\} \exp\left\{\frac{\im}{\hbar}{ \bar{S}}\right\},\texttt{ for }\vp(X)=1, \ gh(X)=-1, \eeq
with the supercommutator $[\ ,\ ]$ and some functional $X=X(\Phi ,\Phi
^{\ast })$, whose form controls the choice of a Lagrangian surface in $%
\Pi T^{\ast }\mathcal{M}$, on which the restriction of the Hessian for $%
S_{X} $ should be non-degenerate. Choosing $X=\Psi (\Phi )$ as the gauge
fermion (e.g., $\Psi (\Phi )=\overline{C}{}^{\alpha }\chi _{\alpha }(A,B)$
for irreducible theories with an admissible gauge $\chi _{\alpha }(A,B)=0$),
one makes the quantum action $S_{\Psi }$ non-degenerate in the configuration
space $\mathcal{M}$,%
\begin{equation}\label{Spsi}
\exp \left\{ \frac{\text{i}}{\hbar }{S_{\Psi }}\right\} =\exp \left\{ \frac{%
\delta \Psi }{\delta \Phi ^{A}}\frac{\delta \ }{\delta \Phi _{A}^{\ast }}%
\right\} \exp \left\{ \frac{\text{i}}{\hbar }{\bar{S}}\right\}
\Longleftrightarrow S_{\Psi }(\Phi ,\Phi ^{\ast })\ =\ {\bar{S}}\big(\Phi
,\,\Phi ^{\ast }+{\textstyle\frac{\delta \Psi }{\delta \Phi }}\big)\ .
\end{equation}

By construction, the action $S_{\Psi }$ satisfies the master equation (\ref{MEBV}) due to the supercommutativity of the operators $\exp \left\{
-[\Delta ,\Psi ]\right\} $ and $\Delta $, namely,
\begin{equation}\label{qmespsi}
\left[ \Delta ,\,\exp \left\{ -[\Delta ,\Psi ]\right\} \right] = 0
\Longrightarrow \Delta \exp \left\{ \frac{\text{i}}{\hbar }{S_{\Psi }}%
\right\} = {0},
\end{equation}%
and is used to construct the path integral and the generating
functionals of Green's functions in the field-antifield
formalism~\cite{BV1}, \cite{BV2}. The
generating functionals of the usual, $Z=Z(J,\Phi ^{\ast })$, and connected, $%
W=W(J,\Phi ^{\ast })$, Green functions extended by external [those
which do not enter the integration measure in (\ref{ZWBV})]
antifields in the BV formalism~\cite{BV1}, \cite{BV2} can be
presented as
\beq \label{ZWBV} \exp\left\{\frac{\im}{\hbar}W\right\}\ =\ Z \ = \ \int\!D\Phi\ \exp
\Big\{\frac{\im}{\hbar} \big(S_\Psi(\Phi,\Phi^*)+ J_A\Phi^A\big)\Big\}\ ,
\eeq
with sources $J_{A}$ ($\varepsilon (J_{A})=\varepsilon _{A}$),
whereas the effective action $\Gamma =\Gamma (\Phi ,\Phi ^{\ast
})$ is determined by the Legendre transformation of $W$ with
respect to $J_{A}$,
%%%%%%%%%%%%%%
\beq \label{EABV} \Gamma (\Phi,\,\Phi^*) \=
W(J,\Phi^*) - J_A\Phi^A, \qquad\textrm{with}\quad \Phi^A =
\frac{\delta W}{\delta J_A}, \qquad
\frac{\delta\Gamma}{\delta\Phi^A}=-J_A\ . \eeq
%%%%%%%%%%%%%
The standard
properties of the above generating functionals are inherited from
the gauge invariance of the classical action, transformed into the
BRST invariance,  being an invariance under global $N=1$
supersymmetry transformations in the extended configuration space $\mathcal{M%
}$,
\beq\label{BRSTc}
\delta_\mu \Phi^A \ =\ (\Phi^A, S_\Psi)\mu,  \quad \delta_\mu \Phi^*_A \ =\ 0,
\eeq
with a constant anticommuting parameter $\mu $.

First, the integrand in Eq. (\ref{ZWBV}) for $Z_{\Psi }\equiv Z(0,\Phi
^{\ast })$ is invariant with respect to the transformations (\ref{BRSTc}).

Second, the vacuum functional $Z_{\Psi }$ is independent with
respect to a variation of the gauge condition, $\Psi \rightarrow
\Psi +\delta \Psi $, if one makes in $Z_{\Psi +\delta \Psi}$ the
change of variables
\beq\label{BRSTg}
\Phi^A \ \to\ \Phi'{}^A = \Phi^A +  (\Phi^A, S_\Psi)\mu(\Phi),  \texttt{ with }  \Phi^{\prime \ast}_A \ =\ \Phi^{\ast}_A,
\eeq
referred to as field-dependent (i.e., gauged) BRST
transformations,\footnote{Despite the term ``gauged'', the
parameter $\mu(\Phi)$ should be considered as an odd-valued functional,
i.e., not as an arbitrary space-time function, such as the gauge
parameter $\xi^{\alpha_0}$.} now with an arbitrary anticommuting
$\mu (\Phi )$, $\mu ^{2}(\Phi )=0$, being, however, infinitesimal,
$\mu (\Phi )=\frac{\text{i}}{\hbar }\delta \Psi $. Indeed, in this
case we have $Z_{\Psi +\delta \Psi }=Z_{\Psi }+o(\delta \Psi )$.

The next consequence of the transformations (\ref{BRSTc}), based on the
equivalence theorem \cite{FradkinTyutin}, is the presence of the Ward
identities for $Z,W,\Gamma $, namely,%
\beq\label{WIBV}
 J_A \frac{\delta Z}{\delta\Phi^*_A} \ = \ 0,\qquad J_A \frac{\delta W}{\delta\Phi^*_A} \ = \ 0,\qquad (\Gamma, \Gamma)\ =\ 0.
\eeq
Finally, the study of gauge dependence for the generating
functionals of Green's functions $Z,W,\Gamma $ leads to the
following variations \cite{VLT, llr, lrr} under the change of the
gauge condition $\Psi \rightarrow \Psi
+\delta \Psi $:%
\beq && \hspace{-1em} \delta Z(J,\Phi^*) \ = \
\frac{\im}{\hbar}J_A\frac{\delta}{\delta\Phi^*_A}\,
 \delta\Psi
\big(\sfrac{\hbar}{\im}\sfrac{\delta}{\delta J}\big)Z(J,\Phi^*)\label{varZBV},
\\
\label{varWBV} \hspace{-1em} && \delta W(J,\Phi^*)\ =\
J_A\frac{\delta }{\delta\Phi^*_A} \delta\Psi \big(\sfrac{\delta
W}{\delta J}+\sfrac{\hbar}{\im}\sfrac{\delta}{\delta J}\big) \\
\label{varGBV} &&\hspace{-1em} \delta\Gamma(\Phi,\Phi^*) \ =\
-(\Gamma,\langle\delta\Psi\rangle)\texttt{ for
}\langle\delta\Psi\rangle\= \delta\Psi({\widehat\Phi})\cdot 1, \
{\widehat\Phi}{}^A\ = \Phi^A + \im\hbar\,(\Gamma^{''-1})^{AB}
\frac{\delta_l}{\delta\Phi^B}, \eeq with the matrix $(\Gamma
^{^{\prime \prime }-1})$ being reciprocal to the Hessian $(\Gamma
^{^{\prime \prime }})$, with the elements \beq \label{invGBV}
(\Gamma^{''})_{AB}\ =\ \frac{\delta_l}{\delta\Phi^A}
\Big(\frac{\delta\Gamma}{\delta\Phi^B}\Big)\ : \quad
(\Gamma^{''-1})^{AC}(\Gamma^{''})_{CB}=\delta^A_{\ B}\ . \eeq

The above local representation for $\delta \Gamma $ can be
rewritten with
the use of differential consequences of the Legendre transformation (\ref%
{EABV}) in a non-local form:
\beq
\delta\Gamma &=&  \frac{\delta\Gamma}{\delta\Phi^A}\Big[-\frac{\delta}{\delta\Phi^*_A}\ +\
(-1)^{\vp_B(\vp_A+1)} (\Gamma^{''-1})^{BC}\Big(\frac{\delta_{\it
l}}{\delta\Phi^C}\frac {\delta
\Gamma}{\delta\Phi^{*}_{A}}\Big)\frac{\delta_{\it l}
}{\delta\Phi^B}\Big]\langle\delta\Psi\rangle\ . \label{varGammaBV}
\eeq
Indeed, in order to derive (\ref{varZBV}) we make the change of variables (%
\ref{BRSTg}) with $\mu (\Phi )=-\frac{\text{i}}{\hbar }\delta \Psi $ in the
functional $Z(J,\Phi ^{\ast })\equiv Z_{\Psi }(J,\Phi ^{\ast })$,
constructed with respect to the action $S_{\Psi }$, then, extracting the
functional $Z_{\Psi +\delta \Psi }(J,\Phi ^{\ast })$, we obtain, with
accuracy up to the first order in $\delta \Psi $,
\begin{eqnarray}
Z_{\Psi }(J,\Phi ^{\ast }) &=&\int \!D\Phi ^{\prime }\ \exp \Big\{\frac{%
\text{i}}{\hbar }\big(S_{\Psi }(\Phi ^{\prime },\Phi ^{\prime \ast
})+J_{A}\Phi ^{\prime }{}^{A}\big)\Big\}  \notag \\
&=&\int \!D\Phi \ \exp \Big\{\frac{\text{i}}{\hbar }\big(S_{\Psi +\delta
\Psi }(\Phi ,\Phi ^{\ast })+J_{A}\Phi ^{A}\big)%
\Big\}\Big(1-\Big(\frac{\text{i}}{\hbar }\Big) ^2%
J_{A}\frac{\delta S_{\Psi }}{\delta \Phi _{A}^{\ast }}\delta \Psi(\Phi) \Big)  \notag \\
&=&Z_{\Psi +\delta \Psi }(J,\Phi ^{\ast })-\frac{\text{i}}{\hbar }J_{A}\frac{%
\delta }{\delta \Phi _{A}^{\ast }}\,\delta \Psi \big({\textstyle\frac{\hbar
}{\text{i}}}{\textstyle\frac{\delta }{\delta J}}\big)Z_{\Psi }(J,\Phi ^{\ast
})\,,  \label{dZderiv}
\end{eqnarray}%
where the final line has been derived  using the differentiation
of the functional integral with respect to the sources $J$ and
external antifields $\Phi^*$.

From the final variations of $Z,W,\Gamma $, we can see, on the
extremals, that for $Z,W$ with $J_{A}=0$ and equivalently for $\Gamma $ with $\frac{%
\delta \Gamma }{\delta \Phi ^{A}}=0$, the corresponding variations
given by Eqs. (\ref{varZBV}), (\ref{varWBV}) and
(\ref{varGammaBV}) are vanishing. This result for $Z,W$ is identical
with that for the vacuum functionals $Z_{\Psi },W_{\Psi
}=\frac{\hbar }{\text{i}}\,\mbox{ln}\,Z_{\Psi }$. The same results
are valid for renormalizable generating  functionals $Z_R,W_R,
\Gamma_R$ with an appropriate gauge-invariant regularization
respecting the Ward identities (\ref{WIBV}) and their differential
consequences.

Due to Gribov's \cite{Gribov} and, in general, Singer's
\cite{Singer} results, we notice that the above-listed BV
quantization rules correctly describe physics within the
functional integral technique in the perturbative way only for
Abelian gauge theories in any gauges and non-Abelian gauge
theories in a connected domain of the configuration space where
the Faddeev--Popov operator (for continuous gauges with space-time
derivatives) for a theory in question has positive eigenvalues.

\subsection{Gauged (Field-dependent) BRST Symmetry Transformations}

\label{gdepBRSTtrans}

Because of the crucial importance of gauged BRST transformations (\ref{BRSTg}%
), we now consider them in detail, assuming that, in general, an
infinitesimal value of the odd-valued parameter $\mu $ can be changed to
a finite nilpotent one, $\Lambda (\Phi ,\Phi ^{\ast })$,
$\Lambda^2 = 0$, being dependent, as a functional, on the entire
set of fields $\Phi ^{A}$ and antifields $\Phi _{A}^{\ast }$
(however, not on the
space-time coordinates in a manifest form) as follows:%
\beq\label{BRSTgen}
\Phi'{}^A = \Phi^A +  (\Phi^A, S_\Psi)\Lambda(\Phi,\Phi^*)  \Longrightarrow  \delta\Phi^A \ =\ S^A_\Psi \Lambda(\Phi,\Phi^*),\texttt{ for }S^A_\Psi\equiv \frac{\delta S_\Psi}{\delta\Phi^*_A}.
\eeq
The corresponding \emph{extended} (due to the antifields) \emph{Slavnov variation}, $%
s_{e}F(\Phi ,\Phi ^{\ast })=\frac{\delta F}{\delta \Phi ^{A}}S_{\Psi }^{A}$
, of an arbitrary functional $F(\Phi ,\Phi ^{\ast })$ generally fails to
be nilpotent,
\begin{equation}
s_{e}^{2}F(\Phi ,\Phi ^{\ast })=\frac{\delta F}{\delta \Phi ^{A}}S_{\Psi
,B}^{A}S_{\Psi }^{B}=\frac{\delta F}{\delta \Phi ^{A}}\big(S_{\Psi
},_{B}S_{\Psi }^{AB}-i\hbar \Delta S_{\Psi }^{A}\big)(-1)^{\varepsilon
_{A}}\neq 0,  \label{Slavnovvar}
\end{equation}%
even for a local action functional, when $\Delta S_{\Psi }^{A}\sim
\delta (0) $.\footnote{The fact that the odd operator $s_{e}$ is
not nilpotent implies that one cannot restore a finite BRST flow
(transformations) in $\Pi T^{\ast }\mathcal{M}$ following the
Frobenius theorem, because the odd-valued vector field
${s}_e\left(\Phi,\Phi^*\right)=\frac{\overleftarrow{\delta}}{\delta \Phi^A}
({s}_e\Phi^A)$ does not have to be nilpotent.} In spite of the
result (\ref{Slavnovvar}), i.e., that $ ({s}_e)^2\ne 0$, the
observation  that for any  constant odd scalar parameters
$\Lambda_1, \Lambda_2$ with  $\mathrm{gh}(\Lambda_1)=
\mathrm{gh}(\Lambda_2)$ there exists a real number $a$ such that $
\Lambda_2=a\Lambda_1$ implies that the right transformations $ G
g(\Lambda)=(1+ {s}_e G \Lambda) $ acting on any functional
$G=G(\Phi,\Phi^*)$ form an Abelian one-parametric supergroup,
since $g(\Lambda_1)g(\Lambda_2) = g(\Lambda_1+\Lambda_2) $ for any
odd $\Lambda_i$, $i=1,2$, due to the fact that $\Lambda_1 \cdot
\Lambda_2 = a\Lambda_1^2=0$.

The usual \emph{Slavnov variation} $sF(\Phi )$ acting on a
functional in the configuration space $F(\Phi )=F(\Phi,0)$ and
determined at the classical level (in the tree approximation) for
$S_{\Psi }=\sum_{k\geq 0}S_{\Psi }^{(k)}$ is not nilpotent,
compared to first-rank gauge theories, including the Yang--Mills
theory \cite{books},
\begin{eqnarray}
&&sF(\Phi )\ =\ \frac{\delta F}{\delta \Phi ^{A}}S^{(0)}{}_{\Psi }^{A}(\Phi
,0),\qquad sF(\Phi )\ =\ s_{e}F(\Phi ,\Phi ^{\ast })\big\vert_{\Phi ^{\ast
}=0},  \label{Slavnovvarr} \\
&&s^{2}F(\Phi )=\frac{\delta F}{\delta \Phi ^{A}}S^{(0)}{}_{\Psi
,B}^{A}S^{(0)}{}_{\Psi }^{B}\big\vert_{\Phi ^{\ast }=0}=\frac{\delta F}{%
\delta \Phi ^{A}}S_{\Psi }^{(0)},_{B}S^{(0)}{}_{\Psi }^{AB}(-1)^{\varepsilon
_{A}}\big\vert_{\Phi ^{\ast }=0}\neq 0.
\end{eqnarray}%
This is explained by an open algebra, described here by the terms $%
S^{(0)}{}_{\Psi }^{AB}|_{\Phi ^{\ast }=0}$, which emerges in the
general Lie bracket for the generators of gauge transformations, $R_{\alpha _{0}}^{i}(A)$%
,%
\begin{equation}
R_{\alpha _{0}}^{i},_{j}(A)R_{\beta _{0}}^{j}(A)-R_{\beta
_{0}}^{i},_{j}(A)R_{\alpha _{0}}^{j}(A)=-R_{\gamma _{0}}^{i}(A)F_{\alpha
_{0}\beta _{0}}^{\gamma _{0}}(A)+S_{0,j}(A)M_{\alpha _{0}\beta _{0}}^{ij}(A),
\label{commgtr}
\end{equation}%
in the form of the coefficients $M_{\alpha _{0}\beta _{0}}^{ij}(A)$ at the
extremals, which, together with the functions $F_{\alpha _{0}\beta
_{0}}^{\gamma _{0}}(A)$, satisfy the properties of generalized antisymmetry%
\begin{equation*}
\lbrack F_{\alpha _{0}\beta _{0}}^{\gamma _{0}},\,M_{\alpha _{0}\beta
_{0}}^{ij}]\ =\ -(-1)^{\varepsilon _{\alpha _{0}}\varepsilon _{\beta
_{0}}}[F_{\beta _{0}\alpha _{0}}^{\gamma _{0}},\,M_{\beta _{0}\alpha
_{0}}^{ij}],\qquad M_{\alpha _{0}\beta _{0}}^{ij}=-(-1)^{\varepsilon
_{i}\varepsilon _{j}}M_{\alpha _{0}\beta _{0}}^{ji}.
\end{equation*}

Let us now calculate the Jacobian of the change of variables
generated by the finite gauged BRST transformations
(\ref{BRSTgen}), namely,
\begin{equation}\label{sdet}
\hspace{-0.05em}\mathrm{Sdet}\left\Vert \frac{\delta \Phi ^{\prime }{}^{A}}{\delta \Phi
{}^{B}}\right\Vert \ =\ \exp \left\{ \mathrm{Str}\,\mathrm{ln}\left( \delta
_{B}^{A}+\frac{\delta (S_{\Psi }^{A}\Lambda )}{\delta \Phi {}^{B}}\right)
\right\} \ =\ \exp \bigg\{-\mathrm{Str}\sum_{n=1}\frac{(-1)^{n}}{n}%
\bigg(\frac{\delta (S_{\Psi }^{A}\Lambda )}{\delta \Phi {}^{B}}\bigg)^{n}%
\bigg\},
\end{equation}
where
\begin{equation}\label{STr}
\mathrm{Str}\bigg(\frac{\delta (S_{\Psi }^{A}\Lambda )}{\delta \Phi {}^{B}}%
\bigg)^{n}\ =\ \frac{\delta (S_{\Psi }^{A}\Lambda )}{\delta \Phi {}^{B_{1}}}%
\frac{\delta (S_{\Psi }^{B_{1}}\Lambda )}{\delta \Phi {}^{B_{2}}}\cdots
\frac{\delta (S_{\Psi }^{B_{n-1}}\Lambda )}{\delta \Phi {}^{A}}%
(-1)^{\varepsilon _{A}}.
\end{equation}%
Explicitly, the supermatrix $\Vert (S_{\Psi }^{A}\Lambda ),_{B}\Vert $  in (\ref{STr}) can
be presented as the sum of two terms:
\begin{equation}\label{P+Q}
(S_{\Psi }^{A}\Lambda ),_{B}\ =\ S_{\Psi }^{A},_{B}\Lambda
(-1)^{\varepsilon _{B}}+S_{\Psi }^{A}\Lambda ,_{B}\ \equiv
P_{B}^{A}+Q_{B}^{A},\texttt{ for }\Lambda_{B}\equiv\Lambda ,_{B},
\end{equation}%
such that only the first supermatrix is nilpotent, $S_{\Psi
}^{A},_{B}\Lambda S_{\Psi }^{B},_{C}\Lambda =P_{B}^{A}P_{C}^{B}=0$, and
furthermore
\begin{equation}\label{nilp}
P_{B}^{A}Q_{C_{1}}^{B}\cdots
Q_{C_{k}}^{C_{k-1}}P_{D}^{C_{k}}Q_{D_{1}}^{D}\cdots Q_{D_{l}}^{D_{l-1}}\ =\
0,
\end{equation}%
for any natural numbers $k,l$.

Using the property of supercommutativity for arbitrary even supermatrices $%
F,G$ under the symbol ``$\mathrm{Str}$'',
$\mathrm{Str}(FG)=\mathrm{Str}(GF)$, we obtain from Eqs.
(\ref{STr}), (\ref{P+Q})  the representation
\begin{equation}\label{str1}
\mathrm{Str}\bigg(P_{B}^{A}+Q_{B}^{A}\bigg)^{n}\ =\ \sum_{k=n-1}^{n}C_{n}^{k}%
\mathrm{Str}\left( (P^{n-k})_{C}^{A}(Q^{k})_{B}^{C}\right) \ =\ n\,\mathrm{Str}%
\left( P_{C}^{A}(Q^{n-1})_{B}^{C}\right) +\mathrm{Str}\left( Q^{n}\right)
_{B}^{A}\ ,
\end{equation}%
with the number of combinations being $C_{n}^{k}=\frac{n!}{k!\left(
n-k\right) !}$.

Consequently, we have
\begin{eqnarray}
&&-\sum_{n=1}^{\infty }\frac{(-1)^{n}}{n}\mathrm{Str}\bigg(\frac{\delta
(S_{\Psi }^{A}\Lambda )}{\delta \Phi {}^{B}}\bigg)^{n}\ =\
\sum_{n=1}^{\infty }\frac{(-1)^{n}}{n}(\Lambda _{C}S_{\Psi }^{C})^{n}  \notag
 \\
&&-\sum_{n=2}^{\infty }{(-1)^{n}}(\Lambda _{C}S_{\Psi }^{C})^{n-2}\Lambda
_{A}S_{\Psi }^{A},{}_{B}S_{\Psi }^{B}\Lambda +S_{\Psi }^{A},{}_{A}\Lambda
\notag \\
&=&\ \sum_{n=1}^{\infty }\frac{(-1)^{n}}{n}(s_{e}\Lambda
)^{n}-\sum_{n=2}^{\infty }{(-1)^{n}}(s_{e}\Lambda )^{n-2}\Lambda _{A}\bigl(%
s_{e}S_{\Psi }^{A}\bigr)\Lambda +\bigl(\Delta S_{\Psi }\bigr)\Lambda  \notag
\\
&=&\ -\,\mbox{ln}\,\big(1+s_{e}\Lambda \big)- %
\sum_{n=1}^{\infty }{(-1)^{n-1}}(s_{e}\Lambda )^{n-1}%
\Lambda _{A}\bigl(s_{e}S_{\Psi }^{A}\bigr)\Lambda +\bigl(\Delta S_{\Psi }%
\bigr)\Lambda  \notag \\
&=&\ -\,\mbox{ln}\,\big(1+s_{e}\Lambda \big)-%
\big(1+s_{e}\Lambda \big)^{-1}\Lambda _{A}\bigl(s_{e}S_{\Psi }^{A}%
\bigr)\Lambda +\bigl(\Delta S_{\Psi }\bigr)\Lambda .\label{strMinter}
\end{eqnarray}%
As a result, we obtain the Jacobian for general gauged
(field-dependent) BRST transformations:
\begin{eqnarray}
&&\mathrm{Sdet}\left\Vert \frac{\delta \Phi ^{\prime }{}^{A}}{\delta \Phi
{}^{B}}\right\Vert \ =\ \big(1+s_{e}\Lambda \big)^{-1}  \exp \left\{ -%
\big(1+s_{e}\Lambda \big)^{-1}\Lambda _{A}\bigl(%
s_{e}S_{\Psi }^{A}\bigr)\Lambda +\bigl(\Delta S_{\Psi }\bigr)\Lambda \right\}
\notag   \\
&&\ =\ \big(1+s_{e}\Lambda \big)^{-1} \Big\{1- \big(%
1+s_{e}\Lambda \big)^{-1}\Lambda _{A}\bigl(s_{e}S_{\Psi }^{A}\bigr)%
\Lambda +\bigl(\Delta S_{\Psi }\bigr)\Lambda \Big\} \nonumber \\
&&\ =\ \big(1+s_{e}\Lambda \big)^{-1} \Big\{1 +
\overleftarrow{s}_{e}\Lambda\Bigr\}\Big\{1 +  \bigl(\Delta S_{\Psi
}\bigr)\Lambda\Bigr\}, \label{jacobianres}\footnotemark
\end{eqnarray}%
\footnotetext{Let us emphasize
that the superdeterminant (\ref{sdet}) for vanishing antifields
$\Phi^*$ was calculated also in \cite{elr}; however, any explicit
calculation is absent. At the same time, the first version of the
present work appeared as arXiv:1312.2092v1[hep-th] earlier than
the above paper \cite{elr}, which appeared as
arXiv:1312.2802v1[hep-th]. Note that in the first version of
\cite{elr} this superdeterminant was calculated in (3.10)
incorrectly. A correct calculation algorithm, including a
representation of the superdeterminant as a series in
(\ref{strMinter}), was given in the 3rd line of (2.33) in
arXiv:1312.2092v1[hep-th]. A small error in the representation of
the second series in the 3rd line, which appeared starting from
the 4th line, was corrected in arXiv:1312.2092v3[hep-th]. Turning
to \cite{elr}, we have to pay attention to the fact that the
introduction of ``finite supersymmetric two-parametric
transformations'' linear in the anticommuting parameters $\xi_a,
a=1, 2$ in (2.3), (2.4) leads to the fact that the set of
field-theoretic models which should be invariant under such
transformations (2.3), (2.4) is  empty because a realistic gauge
field theory (e.g., Yang--Mills theories) should be invariant with
respect to polynomial (in $\xi_a$) transformations, instead of
(2.3); see \cite{MRnew} for details.}where $G\overleftarrow{s}_{e}\equiv s_{e}G$, for any
$G=G\left(\Phi, \Phi^\ast \right)$ and account has been taken of
the identity $\exp \{-a\Lambda \}=1-a\Lambda $, in view of
$\Lambda ^{2}=0$.

Formula (\ref{jacobianres}) is a natural extension of the result
\cite{ll1}, obtained for Yang--Mills theories, for which $s_e$
coincides with $s$,
\begin{equation}
\Big(\Delta S_{\Psi }\ =\ 0,\qquad S_{\Psi }^{A},_{B}S_{\Psi }^{B}\ =\ s^2 =\
0\Big) \Longrightarrow \mathrm{Sdet}\left\Vert \frac{\delta \Phi ^{\prime }{}^{A}}{%
\delta \Phi {}^{B}}\right\Vert \ =\ \big(1+s\Lambda \big)^{-1}, \label{closedJac}
\end{equation}%
and which has also been taken into account as regards the
independence of the generators $S_{\Psi }^{A}(\Phi )$ of BRST
transformations on the antifields $\Phi _{A}^{\ast }$, whereas
the odd-valued functional $\Lambda$ should now be regarded as a
field-dependent one, $\Lambda=\Lambda(\Phi)$. The representation
(\ref{closedJac}) is now valid for a gauge theory of rank 1 with a
closed algebra, and with the additional requirement for the
generators $S_{\Psi }^{A}$ to be divergentless: $\Delta S_{\Psi }
= S_{\Psi }^{A},_{A} = 0$.

For the functional integral
\begin{equation*}
\mathcal{G}(\Phi ^{\ast })\ =\ \int \!D\Phi \ \exp \Big\{\frac{\text{i}}{%
\hbar }Q(\Phi ,\Phi ^{\ast })\Big\},
\end{equation*}%
the change of variables (\ref{BRSTgen}) leads to the representation
\begin{eqnarray}
\mathcal{G}(\Phi ^{\ast }) &=&\int \!D\Phi ^{\prime }\ \exp \Big\{\frac{%
\text{i}}{\hbar }Q(\Phi ^{\prime },\Phi ^{\ast })\Big\}=  \int \!D\Phi \ \exp \Big\{\frac{\text{i}}{\hbar }\Big(Q(\Phi ,\Phi ^{\ast
})\notag \\
&+&s_{e}Q\Lambda (\Phi ,\Phi ^{\ast })-i\hbar \bigl(\Delta S_{\Psi }\bigr)%
\Lambda -i\hbar \,\mbox{ln}\,\left[\big(1+s_{e}\Lambda \big)^{-1}\big(1 +  \overleftarrow{s}_{e}\Lambda\big)\right]   \Big)\Big\}
,  \label{genY}
\end{eqnarray}%
which is different from the similar result \cite{ll1} for
the Yang--Mills theory due to the terms
$\big(1 +  \overleftarrow{s}_{e}\Lambda\big)$
in the final line and the presence of $i\hbar %
\bigl(\Delta S_{\Psi }\bigr)\Lambda$.

A repeated application of gauged BRST transformations with the same gauged
parameter $\Lambda $ is not nilpotent due to the equality
\beq
\label{notnilp}
\delta_\Lambda\left(\delta_\Lambda F(\Phi,\Phi^*)\right)  \ =\ \delta_\Lambda\left(s_e F(\Phi,\Phi^*)\Lambda\right) \ =\ - s^2_e F(\Phi,\Phi^*)\Lambda^2 +  s_e F(\Phi,\Phi^*)s_e(\Lambda)\Lambda,
\eeq
and due to the vanishing commutator $[\delta _{\mu_{1}},\delta
_{\mu _{2}}]F = 2(s^2_e F) \mu _{1}\mu _{2} = 0$ of global BRST transformations with constant
parameters $\mu _{1},\mu _{2}$, such that $\mu _{1}= a\cdot\mu _{2}$, as shown by Eq. (\ref{Slavnovvar}) for $%
s_{e}^{2}F\neq 0$ and the subsequent relations. Of course, for a constant $\Lambda $ (i.e., $\Lambda =\mu $%
) the nilpotency in Eq. (\ref{notnilp}) is restored; however, as
compared with the Yang--Mills theory we have the standard
expression for the Jacobian:
\begin{equation*}
\mathrm{Sdet}\left\Vert \frac{\delta \Phi ^{\prime }{}^{A}}{\delta \Phi
{}^{B}}\right\Vert \ =\exp \bigl\{\Delta S_{\Psi }\mu \bigr\}.
\end{equation*}

Let us now examine the generating functionals $Z_{\Psi }(J,\Phi ^{\ast })$, $%
Z_{\Psi +\overline{\Delta }\Psi }(J,\Phi ^{\ast })$ in Eq. (\ref%
{dZderiv}) for the same gauge theory, however, given by different
(not necessarily related to each other by small variations) gauges
described by
the gauge fermions $\Psi (\Phi )$, $[\Psi +\overline{\Delta }\Psi ](\Phi )$%
, which differ by a Grassmann-odd  functional, $\overline{\Delta
}\Psi (\Phi )$, subject to the conditions $(\varepsilon
,gh)\overline{\Delta }\Psi (\Phi )\ =\ (1,-1)$.

After the change of variables (\ref{BRSTgen}), due to the quantum
master equation (\ref{qmespsi}) for the action $S_{\Psi }$, we
obtain the generating functional $Z_{\Psi }(J,\Phi ^{\ast })$, namely,%
\begin{eqnarray}
Z_{\Psi }(J,\Phi ^{\ast }) &=&\int \!D\Phi \ \exp \Big\{\frac{\text{i}}{%
\hbar }\Big(S_{\Psi }+i\hbar \,\mbox{ln}\,\big(1+s_{e}\Lambda \big)  \notag
\label{changZ} \\
&&+i\hbar \big(1+s_{e}\Lambda \big)^{-1}%
\Lambda _{A}\bigl(s_{e}S_{\Psi }^{A}\bigr)\Lambda +J_{A}(\Phi ^{A}+\delta
\Phi ^{A})\Big)\Big\}.
\end{eqnarray}%
In its turn, $Z_{\Psi +\overline{\Delta }\Psi }(J,\Phi ^{\ast })$
corresponding to a finite change of the gauge fermion takes the
form
\begin{eqnarray}
Z_{\Psi +\overline{\Delta }\Psi }(J,\Phi ^{\ast }) &=&\int \!D\Phi \ \exp %
\Big\{\frac{\text{i}}{\hbar }\Big(S_{\Psi }(\Phi ,\Phi ^{\ast })+s_{e}\bigl(%
\overline{\Delta }\Psi (\Phi )\bigr)  \notag  \label{changZ2} \\
&&+\sum_{n\geq 2}\frac{1}{n!}\overline{\Delta }\Psi _{A_{1}}\cdots \overline{%
\Delta }\Psi _{A_{n}}S_{\Psi }^{A_{n}...A_{1}}(\Phi ,\Phi ^{\ast
})+J_{A}\Phi ^{A}\Big)\Big\}.
\end{eqnarray}
Consider a functional equation for an unknown odd-valued functional, $%
\Lambda $, following the requirement of coincidence of the above
representations (\ref{changZ}) and (\ref{changZ2}) for $J_{A}=0$: $Z_{\Psi +\overline{\Delta }\Psi }(0,\Phi ^{\ast })=Z_{\Psi }(0,\Phi ^{\ast })$
\begin{eqnarray}
&&i\hbar \left\{ \,\mbox{ln}\,\big(1+s_{e}\Lambda \big)+\big(1+s_{e}\Lambda \big)^{-1}\Lambda _{A}\bigl(s_{e}S_{\Psi
}^{A}\bigr)\Lambda \right\}= \sum_{n\geq 1}\frac{1}{n!}%
\overline{\Delta }\Psi _{A_{1}}\cdots \overline{\Delta }\Psi _{A_{n}}S_{\Psi
}^{A_{n}...A_{1}}(\Phi ,\Phi ^{\ast }) \notag   \\
&& \Longleftrightarrow -i\hbar  \,\mbox{ln}\,\Big\{\big(1+s_{e}\Lambda \big)^{-1} \big(1+\overleftarrow{s}_{e}\Lambda \big) \Big\}  = \Big(\exp\Big\{ -[\Delta,\,\overline{\Delta} \Psi] \Big\}-1\Big)S_\Psi .\label{eqlambdapsi}
\end{eqnarray}%
Having in mind the fact that for an infinitesimal
$\overline{\Delta }\Psi =\delta \Psi $, with accuracy up to the
first order in $\delta \Psi $
from Eq. (\ref{eqlambdapsi}),  we have a linearized (with respect to $\Lambda $ and $%
\Lambda_{A}$) and easily solved equation,
\beq\label{eqlambdapsilin} i\hbar s_e \Lambda \ =\ s_e
{\delta}\Psi(\Phi) \Longrightarrow \Lambda = - \frac{i}{\hbar}
{\delta}\Psi\texttt{ and }\Lambda \ = \ \Lambda(\Phi). \eeq This
fact has been used to verify the gauge-independence property
$Z_{\Psi
}=Z_{\Psi +\delta \Psi }$ for the vacuum functional $Z_{\Psi }$ in Section~%
\ref{generals}.

Therefore, we hope that the highly non-linear equation
(\ref{eqlambdapsi}), which provides a compensation for a finite
change of the gauge Fermion in $Z_{\Psi }$ by means of the
Jacobian for the change of variables generated by the gauged BRST
transformations (\ref{BRSTgen}), also has a solution, which should
be of the form
\begin{equation}\label{solcompeq}
\Lambda\left(\Phi,\Phi ^{\ast }|\overline{\Delta} \Psi\right)  = \Lambda\left(\overline{\Delta} \Psi\right).
\end{equation}
We analyze a justification of this representation in
Appendix~\ref{AppA}.

Using the above result, we argue that that for any finite change
of the gauge $\overline{\Delta}\Psi$ there exists a gauged
(field-dependent) BRST transformation (\ref{BRSTg}) with an odd-valued
functional $\Lambda\left(\overline{\Delta} \Psi\right)$
in (\ref{solcompeq}) such that, due
to the equivalence theorem \cite{FradkinTyutin, Tyutin}, there is
a coincidence of the two representations (%
\ref{changZ}) and (\ref{changZ2}), which is also valid
for the vacuum functional:
\begin{equation}\label{equival}
Z_{\Psi +\overline{\Delta }\Psi }(0,\Phi ^{\ast })=Z_{\Psi }(0,\Phi ^{\ast
}).
\end{equation}%
This is the main result of this section, which we use in the study
of the gauge-(in)dependence problem for a theory with BRST
symmetry breaking terms.

\section{Gauge Dependence for Generating Functionals with Broken BRST
Symmetry}
\label{gengd}

\noindent

Let us turn to the problem of gauge dependence for a gauge theory
determined by Eqs. (\ref{GIClassA}), (\ref{redth}), (\ref{redth1}) with a
quantum action $S_{\Psi }(\Phi ,\Phi ^{\ast })$ additively extended
along the lines of our previous study \cite{llr}, \cite{lrr} by a soft
BRST breaking term $M(\Phi ,\Phi ^{\ast })$ defined in a gauge $\Psi
(\Phi )$ up to an action $S(\Phi ,\Phi ^{\ast })$ determining the
generating functional of Green's functions, $Z_{M}(J,\Phi ^{\ast })$,
\beq \label{ZSfull}
S\=S_{\Psi}+M,\quad
Z_M(J,\Phi^*)\= \int\!D\Phi\ \exp \Big\{\frac{\im}{\hbar}
\big(S(\Phi,\Phi^*)+ J_A\Phi^A\big)\Big\}\ , \eeq
with the boundary condition
\begin{equation*}
Z_{M}(J,\Phi ^{\ast })\big\vert_{M=0}\ =\ Z(J,\Phi ^{\ast }).
\end{equation*}%
We remind that, at the classical level, since we assume the bosonic
functional $M(\Phi ,\Phi ^{\ast })$ to have a regular decomposition in powers of
$\hbar $, $M(\Phi ,\Phi ^{\ast })=\sum_{n\geq 0}\hbar ^{n}M_{n}(\Phi ,\Phi
^{\ast })$, the condition of a soft breaking of BRST symmetry implies%
\beq
\label{brbrstclas}
(M_0,M_0)\ =\ 0\texttt{ and } m_0,_i R^i_{\alpha_0}\ne 0,\texttt{ for }  m_0(\Phi) = M_0(\Phi,\Phi^*)\big\vert_{\Phi^*=0},
\eeq
whereas in the case of a regularization more general than dimensional-like ones,
the total generating equation for $M(\Phi ,\Phi ^{\ast })$
reads\footnote{One may examine a more general BRST symmetry breaking functional,
not satisfying Eq. (\ref{brbrst}) or Eq. (\ref{brbrstclas}), without changing
the results for the dependence of the effective action (as will be seen
later); however, we will follow the study of \cite{llr}, \cite{lrr}, because
the solution of these equations restricts the rank condition for the Hessian
of $M$ to be no greater than $\dim \mathcal{M}$, and to be such that the
functional integral in (\ref{ZSfull}) is well-defined.}
\cite{lrr}
\beq\label{brbrst}
\Delta \left\{-\frac{i}{\hbar} M \right\} = 0 \Longleftrightarrow \sfrac{1}{2}(M,M) =-i\hbar\Delta M.
\eeq
As a consequence of
Eqs. (\ref{qmespsi}), (\ref{brbrst}), the total action now satisfies%
\beq
\label{CBasEq}
\sfrac{1}{2}(S,S)-\im\hbar\,{\Delta}{S}\=(S,M)\ ,
\eeq
so that, in the classical limit for $S=S_{0}+O(\hbar )$, Eq. (\ref{CBasEq})
implies the equation
\beq \label{CBasEq0}
\sfrac{1}{2}(S_0,S_0)\=(S_0,M_0).
\eeq
The properties of the generating functionals of the usual, $Z_{M}(J,\Phi
^{\ast })$, connected, $W_{M}(J,\Phi ^{\ast })$, ($W_{M}=\frac{\hbar }{i}\,%
\mbox{ln}\,Z_{M}$) and vertex, $\Gamma _{M}(\Phi ,\Phi ^{\ast })$, Green functions,
introduced via the Legendre
transformation of $W_{M}(J,\Phi ^{\ast })$ with respect to the sources $J_{A}$,
\beq
\label{EAq}
\Gamma_M(\Phi,\Phi^*)\ =\ W_M(J,\Phi^*)-
J_A\Phi^A,\quad
\Phi^A=\frac{\delta W_M(J,\Phi^*)}{\delta J_A},
\eeq
have been studied in \cite{llr}, \cite{lrr}. These properties include the Ward
identities and the calculation of variations of all the generating
functionals under a variation of the gauge condition (Grassmann-odd functional), $\Psi (\Phi
)\rightarrow \Psi (\Phi )+\delta \Psi (\Phi )$. The properties were derived
on the basis of fun\-ctional averaging of the  master equations (\ref%
{qmespsi}) for $S_{\Psi }$ in a dimensional-like regularization, as
applied to the local functional $S$ \cite{llr}, and in more general
regularizations \cite{lrr}.

These properties can only be obtained by means of global BRST and
field-dependent (gauged) BRST transformations. There follow the Ward identities for $%
Z_{M}(J,\Phi ^{\ast })$, after the change of variables (\ref{BRSTc})
in the integrand of (\ref{ZSfull}), with account taken of Eqs. (\ref{qmespsi}%
), (\ref{brbrst}),
\begin{eqnarray}
&&Z_{M}(J,\Phi ^{\ast }) = {\int }\!D\Phi \ \exp \Big\{\frac{\text{i}}{%
\hbar }\Big(S(\Phi ,\Phi ^{\prime \ast })+\frac{\delta S}{\delta \Phi ^{A}}%
\frac{\delta S_{\Psi }}{\delta \Phi _{A}^{\ast }}\mu -i\hbar \Delta S_{\Psi
}\mu +J_{A}\big[\Phi ^{A}+\frac{\delta S_{\Psi }}{\delta \Phi _{A}^{\ast }}%
\mu \Big]\Big)\Big\}\   \notag   \\
&& = \ Z_{M}(J,\Phi ^{\ast })+\frac{i}{\hbar }\int \!D\Phi \ \big(J_{A}+M_{A}%
\big)\frac{\delta S_{\Psi }}{\delta \Phi _{A}^{\ast }}\exp \Big\{\frac{\text{%
i}}{\hbar }\Big(S(\Phi ,\Phi ^{\prime \ast })+J_{A}\Phi ^{A}\Big)\Big\}\mu ,
\notag \\
& & \Longrightarrow \Big(J_{A}+M_{A}\big({\textstyle\frac{\hbar }{\text{i}}}{%
\textstyle\frac{\delta }{\delta J}},\Phi ^{\ast }\big)\Big)\left( \frac{%
\hbar }{\text{i}}\frac{\delta }{\delta \Phi _{A}^{\ast }}\ -\ M^{A\ast }\big(%
{\textstyle\frac{\hbar }{\text{i}}}{\textstyle\frac{\delta }{\delta J}},\Phi
^{\ast }\big)\right) Z_{M}(J,\Phi ^{\ast })=0,\label{wardZM}
\end{eqnarray}%
where the notation
\begin{equation}\label{MAZ}
M_{A}\big({\textstyle\frac{\hbar }{\text{i}}}{\textstyle\frac{\delta }{%
\delta J}},\Phi ^{\ast }\big)\equiv \frac{\delta M(\Phi ,\Phi ^{\ast })}{%
\delta \Phi ^{A}}\Big|_{\Phi \rightarrow \frac{\hbar }{\text{i}}\frac{\delta
}{\delta J}}\qquad \text{and}\qquad M^{A\ast }\big({\textstyle\frac{\hbar }{%
\text{i}}}{\textstyle\frac{\delta }{\delta J}},\Phi ^{\ast }\big)\equiv
\frac{\delta M(\Phi ,\Phi ^{\ast })}{\delta \Phi _{A}^{\ast }}\Big|_{\Phi
\rightarrow \frac{\hbar }{\text{i}}\frac{\delta }{\delta J}}
\end{equation}%
has been used. In case $M=0$, identity (\ref{wardZM}) is reduced to the
usual Ward identity (\ref{WIBV}) for $Z(J,\Phi ^{\ast })$, as well as to the
Ward identities for $W_{M}(J,\Phi ^{\ast })$, $\Gamma _{M}(\Phi ,\Phi ^{\ast
})$, which follow from (\ref{wardZM}),%
\begin{eqnarray}
&&\Big(J_{A}+M_{A}\big({\textstyle\frac{\delta W_{M}}{\delta J}}+{\textstyle%
\frac{\hbar }{\text{i}}}{\textstyle\frac{\delta }{\delta J}},\Phi ^{\ast }%
\big)\Big)\left( \frac{\delta W_{M}(J,\Phi ^{\ast })}{\delta \Phi _{A}^{\ast
}}\ -\ M^{A\ast }\big({\textstyle\frac{\delta W_{M}}{\delta J}}+{\textstyle%
\frac{\hbar }{\text{i}}}{\textstyle\frac{\delta }{\delta J}},\Phi ^{\ast }%
\big)\right) =0,  \label{WIWBV} \\
&&{\textstyle\frac{1}{2}}(\Gamma _{M},\Gamma _{M})\ =\frac{\delta \Gamma _{M}%
}{\delta \Phi ^{A}}{\widehat{M}}^{A\ast }+{\widehat{M}}_{A}\frac{\delta
\Gamma _{M}}{\delta \Phi _{A}^{\ast }}-{\widehat{M}}_{A}{\widehat{M}}^{A\ast
}\ .  \label{WIGammaBV}
\end{eqnarray}%
Here, we have used a notation introduced in \cite{llr}:
\begin{equation}\label{MAG}
{\widehat{M}}_{A}\ \equiv \ \frac{\delta M(\Phi ,\Phi ^{\ast })}{\delta \Phi
^{A}}\Big|_{\Phi \rightarrow \widehat{\Phi }}\qquad \text{and}\qquad {%
\widehat{M}}^{A\ast }\ \equiv \ \frac{\delta M(\Phi ,\Phi ^{\ast })}{\delta
\Phi _{A}^{\ast }}\Big|_{\Phi \rightarrow \widehat{\Phi }}\,,
\end{equation}%
with account taken for the conventions (\ref{varGBV}), (\ref{invGBV}),
adapted to the case of broken BRST symmetry, i.e., according to the  change $\Gamma \to \Gamma_M$. For completeness, note that the
functional\ $\Gamma _{M}$ satisfies the functional integro-differential
equation
\begin{eqnarray}
\exp \Big\{\frac{i}{\hbar }\,\Gamma _{M}(\Phi ,\Phi ^{\ast })\Big\} &=&\int
d\varphi \,\exp \Big\{\frac{i}{\hbar }\Big[S_{\Psi }({\Phi +\hbar ^{\frac{1}{%
2}}\varphi },\Phi ^{\ast })+M(\Phi +\hbar ^{\frac{1}{2}}\varphi ,\Phi ^{\ast
})\,  \notag \\
&&-\,\frac{\delta \Gamma _{M}(\Phi ,\Phi ^{\ast })}{\delta \Phi }\,\hbar ^{%
\frac{1}{2}}\varphi \Big]\Big\}\,,  \label{GMEq-loop}
\end{eqnarray}%
determining the loop expansion $\Gamma _{M}=\sum_{n\geq 0}\hbar ^{n}\Gamma
_{n{}M}$. Thus, the tree-level (zero-loop) and one-loop approximations of (%
\ref{GMEq-loop}) correspond to
\begin{eqnarray}
&&\Gamma _{0{}M}(\Phi ,\Phi ^{\ast })\,=\,S_{\Psi {}0}({\Phi },\Phi ^{\ast
})\,+\,M_{0}(\Phi ,\Phi ^{\ast })\,,  \label{0loop} \\
&&\Gamma _{1{}M}(\Phi ,\Phi ^{\ast })\,=\,S_{\Psi {}1}({\Phi },\Phi ^{\ast
})\,+\,M_{1}(\Phi ,\Phi ^{\ast })-\frac{\text{i}}{2}\,\mbox{ln}\,\mathrm{Sdet%
}\left\Vert (S_{0}^{^{\prime \prime }})_{AB}({\Phi },\Phi ^{\ast
})\right\Vert ,  \label{1loop}
\end{eqnarray}%
so that the tree-level part of the Ward identity (\ref{WIGammaBV}) for $%
\Gamma _{0{}M}$, %%%
\begin{equation*}
{\textstyle\frac{1}{2}}(\Gamma _{0{}M},\Gamma _{0{}M})\ =\frac{\delta
S_{\Psi {}0}}{\delta \Phi ^{A}}{M}_{0}^{A\ast }+{M}_{0{}A}\frac{\delta
S_{\Psi {}0}}{\delta \Phi _{A}^{\ast }}+{M}_{0{}A}{M}_{0}^{A\ast }\
\end{equation*}%
is fulfilled identically, due to the tree-level approximation
to the generating equations (\ref{qmespsi}) for $S_{\Psi {}0}$ and (\ref%
{brbrstclas}) for $M_{0}$.

In order to study the gauge-dependence problem, we examine, first of all,
the representation for $Z_{M}(J,\Phi ^{\ast })$ within the gauge
determined by the gauge functional, ${\Psi +\overline{\Delta }\Psi }$,
similar to Eq. (\ref{changZ2}), but without the use of
field-dependent BRST transformations:
\begin{eqnarray}
Z_{M}(J,\Phi ^{\ast }) &=&\int \!D\Phi \ \exp \Big\{\frac{\text{i}}{\hbar }%
\Big(S_{\Psi }(\Phi ,\Phi ^{\ast })+M(\Phi ,\Phi ^{\ast })+s_{e}\bigl(%
\overline{\Delta }\Psi (\Phi )\bigr)  \notag  \label{varZM} \\
&&+\sum_{n\geq 2}\frac{1}{n!}\overline{\Delta }\Psi _{A_{1}}\cdots \overline{%
\Delta }\Psi _{A_{n}}S_{\Psi }^{A_{n}...A_{1}}(\Phi ,\Phi ^{\ast })+%
\overline{\Delta }M(\Phi ,\Phi ^{\ast })+J_{A}\Phi ^{A}\Big)\Big\},
\end{eqnarray}%
where account has been taken for the fact that the functional $M=M_{\Psi
}(\Phi ,\Phi ^{\ast })$ should have the following representation in the
above gauge, because of the relation of gauged  BRST transformations with functional $ \Lambda\left(\Phi,\Phi ^{\ast }|\overline{\Delta} \Psi\right)$ (\ref{solcompeq}) which should compensate a finite change of the gauge $\overline{\Delta }\Psi $ in $Z_{\Psi}(0,\Phi ^{\ast })$:
\begin{equation}\label{mpsi+}
M_{\Psi +\overline{\Delta }\Psi }(\Phi ,\Phi ^{\ast })\ =\ M_{\Psi }(\Phi
,\Phi ^{\ast })+\overline{\Delta }M(\Phi ,\Phi ^{\ast }).
\end{equation}%
It should be noted that $M_{\Psi +\overline{\Delta }\Psi }$ does not have
the form of a gauge-invariant action, $S_{\Psi +\overline{\Delta }\Psi }$,
as regards the dependence on the variation $\overline{\Delta }M$, despite the
fact that an introduction of the additive term $\overline{\Delta }\Psi $ by
means of the transformation (\ref{qaBV}) applied to the action, $S_{\Psi }$,
\begin{equation}\label{spsi+dp}
\exp \left\{ \frac{\text{i}}{\hbar }{S_{\Psi +\overline{\Delta }\Psi }}%
\right\} =\exp \left\{ -[\Delta ,\overline{\Delta }\Psi ]\right\} \exp
\left\{ \frac{\text{i}}{\hbar }{{S}_{\Psi }}\right\} =\exp \left\{ \frac{%
\text{i}}{\hbar }{{S}_{\Psi }(\Phi ,\Phi ^{\ast }+{\textstyle\frac{\delta
\overline{\Delta }\Psi }{\delta \Phi }})}\right\} ,
\end{equation}%
is a transformation which turns a solution of the soft BRST symmetry breaking
equation (\ref{brbrst}) for $M_{\Psi }$ into another solution, however, not
having the form $M_{\Psi +\overline{\Delta }\Psi }$. This takes place, since
in the case of the functional $M_{\Psi }$, being BRST-non-invariant,
the gauge condition is not determined via a shift of the antifields:
\begin{equation}\label{mpsi+in}
M_{\Psi +\overline{\Delta }\Psi }\neq M_{\Psi }\bigl(\Phi ,\Phi ^{\ast }+{%
\textstyle\frac{\delta \overline{\Delta }\Psi }{\delta \Phi }})\bigr).
\end{equation}%
As we turn to Eq. (\ref{varZM}), let us present the finite change
$\overline{\Delta }Z_{M}(J,\Phi ^{\ast }) =
Z_{M+\overline{\Delta }M, \Psi+\overline{\Delta }\Psi}(J,\Phi ^{\ast })
-Z_{M, \Psi}(J,\Phi ^{\ast })$ in an equivalent form:%
\begin{eqnarray}
\overline{\Delta }Z_{M}(J,\Phi ^{\ast }) &=&\int \!D\Phi \ \Big[\exp \Big\{%
\frac{\text{i}}{\hbar }\Big(s_{e}\bigl(\overline{\Delta }\Psi (\Phi )\bigr)%
+\sum_{n\geq 2}\frac{1}{n!}\overline{\Delta }\Psi _{A_{1}}\cdots \overline{%
\Delta }\Psi _{A_{n}}S_{\Psi }^{A_{n}...A_{1}}(\Phi ,\Phi ^{\ast })  \notag
 \\
&&+\overline{\Delta }M(\Phi ,\Phi ^{\ast })\Big)\Big\}-1\Big]\exp \Big\{%
\frac{\text{i}}{\hbar }\big(S(\Phi ,\Phi ^{\ast })+J_{A}\Phi ^{A}\big)\Big\}  \notag \\
&=&\int \!D\Phi \ \Big[\exp \Big\{\frac{\text{i}}{\hbar }\overline{\Delta }%
M(\Phi ,\Phi ^{\ast })\Big\}\exp \Big\{\frac{\delta \overline{\Delta }\Psi
(\Phi )}{\delta \Phi ^{A}}\Big(\frac{\delta }{\delta \Phi _{A}^{\ast }}-%
\frac{\text{i}}{\hbar }M^{A\ast }(\Phi ,\Phi ^{\ast })\Big)\Big\}-1\Big]
\notag \\
&&\times \exp \Big\{\frac{\text{i}}{\hbar }\big(S(\Phi ,\Phi ^{\ast })+J_{A}\Phi
^{A}\big)\Big\},\label{varZM2}
\end{eqnarray}%
with allowance for the identity
\beq\label{auxexp}
 \Big[\exp \Big\{\frac{\delta \overline{\Delta }\Psi
}{\delta \Phi ^{A}}\frac{\delta S_\Psi}{\delta \Phi _{A}^{\ast }}\Big\}-1\Big]\exp \Big\{\frac{\text{i}}{\hbar }S\Big\}
\ =\
\Big[\exp \Big\{\frac{\delta \overline{\Delta }\Psi
}{\delta \Phi ^{A}}\Big(\frac{\delta }{\delta \Phi _{A}^{\ast }}-%
\frac{\text{i}}{\hbar }M^{A\ast }\Big)\Big\}-1\Big]
\exp \Big\{\frac{\text{i}}{\hbar }S\Big\}.
\eeq
Considering the general term $ \Big\{\frac{\delta \overline{\Delta }\Psi
}{\delta \Phi ^{A}}\Big(\frac{\delta }{\delta \Phi _{A}^{\ast }}-%
\frac{\text{i}}{\hbar }M^{A\ast }\Big)\Big\}^n$, for $n\geq 1$, inside the decomposition (\ref{varZM2}) and
integrating by parts in the path integral, we obtain
 \beq
 &&
 \int \!D\Phi \ \exp \Big\{\textstyle\frac{\text{i}}{\hbar }\overline{\Delta }%
M\Big\}\displaystyle\frac{\delta \overline{\Delta }\Psi
}{\delta \Phi ^{A}}\Big\{ \overline{\Delta }\Psi_B
\textstyle\Big(\frac{\delta }{\delta \Phi _{B}^{\ast }}-%
\frac{\text{i}}{\hbar }M^{B\ast }\Big)\Big\}^{n-1} \Big(\frac{\delta }{\delta \Phi _{A}^{\ast }}-%
\frac{\text{i}}{\hbar }M^{A\ast }\Big)\exp \Big\{\frac{\text{i}}{\hbar }\big(S+J_{A}\Phi
^{A}\big)\Big\}
\notag \\
&&=  \frac{\text{i}}{\hbar }\int \!D\Phi \ \exp \Big\{\textstyle\frac{\text{i}}{\hbar }\overline{\Delta }%
M\Big\}\bigg[\overline{\Delta }%
M_A
\Big\{ \overline{\Delta }\Psi_B
\textstyle\Big(\frac{\delta }{\delta \Phi _{B}^{\ast }}-%
\frac{\text{i}}{\hbar }M^{B\ast }\Big)\Big\}^{n-1} \Big(\frac{\delta }{\delta \Phi _{A}^{\ast }}-%
\frac{\text{i}}{\hbar }M^{A\ast }\Big) \nonumber \\
&& - \Big\{\sum_{k=1}^{n-1}\prod_{l=1}^{k-1}  \overline{\Delta }\Psi_{B_l}
\textstyle\Big(\frac{\delta }{\delta \Phi _{B_l}^{\ast }}-%
\frac{\text{i}}{\hbar }M^{B_l\ast }\Big)\overline{\Delta }\Psi_{B_k}M^{B_k\ast }_{A}\displaystyle\prod\limits_{l=k+1}^{n-1}  \overline{\Delta }\Psi_{B_l}
\textstyle\Big(\frac{\delta }{\delta \Phi _{B_l}^{\ast }}-%
\frac{\text{i}}{\hbar }M^{B_l\ast }\Big) \Big\} \Big(\frac{\delta }{\delta \Phi _{A}^{\ast }}-%
\frac{\text{i}}{\hbar }M^{A\ast }\Big) \nonumber\\
&&  + \Big\{ \overline{\Delta }\Psi_B
\textstyle\Big(\frac{\delta }{\delta \Phi _{B}^{\ast }}-%
\frac{\text{i}}{\hbar }M^{B\ast }\Big)\Big\}^{n-1} \Big(J_A+ M_A\Big)\Big(\frac{\delta }{\delta \Phi _{A}^{\ast }}-%
\frac{\text{i}}{\hbar }M^{A\ast }\Big)\bigg]\overline{\Delta }\Psi \exp \Big\{\frac{\text{i}}{\hbar }\big(S+J_{A}\Phi
^{A}\big)\Big\}, \label{auxfint}
 \eeq
 where account has been taken of the generating equations (\ref{qmespsi})
   for $S_\Psi$ and (\ref{brbrst}) for $M$,
   as well as the notation $M^{B_k\ast }_{A}
   \equiv \frac{\delta}{\delta \Phi^A} (M^{B_k\ast })$
     and the following  properties of the functional $\overline{\Delta }\Psi$:
 \begin{equation*}
 \big(\overline{\Delta }\Psi\big)^2\ \equiv\ 0,\texttt{ and } \overline{\Delta }\Psi_{AB}\Big(\frac{\delta }{\delta \Phi _{B}^{\ast }}-%
\frac{\text{i}}{\hbar }M^{B\ast }\Big)\Big(\frac{\delta }{\delta \Phi _{A}^{\ast }}-%
\frac{\text{i}}{\hbar }M^{A\ast }\Big)\equiv 0.
 \end{equation*}
The variation of the functional $Z_M(J,\Phi ^{\ast })$ can therefore be finally presented
as
\begin{eqnarray}
\overline{\Delta }Z_{M} &=& \int \!D\Phi \ \bigg[\exp \Big\{%
\frac{\text{i}}{\hbar }\overline{\Delta }M\Big\} \sum_{n\geq 0}\frac{1}{n!}\Big\{ \overline{\Delta }\Psi_B
\textstyle\Big(\frac{\delta }{\delta \Phi _{B}^{\ast }}-%
\frac{\text{i}}{\hbar }M^{B\ast }\Big)\Big\}^{n} -1\bigg]\exp \Big\{\frac{\text{i}}{\hbar }\big(S+J_{A}\Phi ^{A}\big)\Big\}    \notag
\\
&=&\frac{\text{i}}{\hbar }\exp \Big\{\frac{\text{i}}{\hbar }\overline{\Delta }M({\textstyle%
\frac{\hbar }{\text{i}}}{\textstyle\frac{\delta }{\delta J}},\Phi ^{\ast })%
\Big\} \bigg[\overline{\Delta }%
M_A
\sum_{n\geq 1}\frac{1}{n!}\Big\{ \overline{\Delta }\Psi_B
\textstyle\Big(\frac{\delta }{\delta \Phi _{B}^{\ast }}-%
\frac{\text{i}}{\hbar }M^{B\ast }\Big)\Big\}^{n-1}  \nonumber \\
&& - \sum_{n\geq 1}\frac{1}{n!}\Big\{\sum_{k=1}^{n-1}\prod_{l=1}^{k-1}  \overline{\Delta }\Psi_{B_l}\textstyle\Big(\frac{\delta }{\delta \Phi _{B_l}^{\ast }}-%
\frac{\text{i}}{\hbar }M^{B_l\ast }\Big)\overline{\Delta }\Psi_{B_k}M^{B_k\ast }_{A}\displaystyle\prod\limits_{l=k+1}^{n-1}  \overline{\Delta }\Psi_{B_l}
\textstyle\Big(\frac{\delta }{\delta \Phi _{B_l}^{\ast }}-%
\frac{\text{i}}{\hbar }M^{B_l\ast }\Big) \Big\}  \nonumber\\
&&  + \sum_{n\geq 1}\frac{1}{n!}\Big\{ \overline{\Delta }\Psi_B
\textstyle\Big(\frac{\delta }{\delta \Phi _{B}^{\ast }}-%
\frac{\text{i}}{\hbar }M^{B\ast }\Big)\Big\}^{n-1} \Big(J_A+ M_A\Big)\bigg]\Big(\frac{\delta }{\delta \Phi _{A}^{\ast }}-%
\frac{\text{i}}{\hbar }M^{A\ast }\Big)\overline{\Delta }\Psi({\textstyle%
\frac{\hbar }{\text{i}}}{\textstyle\frac{\delta }{\delta J}})
Z_{M} \nonumber\\
&& +  \bigg[\exp \Big\{\frac{\text{i}}{\hbar }\overline{\Delta }M({\textstyle%
\frac{\hbar }{\text{i}}}{\textstyle\frac{\delta }{\delta J}},\Phi ^{\ast })%
\Big\}-1\bigg]Z_{M},\label{varZMf}
\end{eqnarray}%
where the arguments ${\textstyle%
\frac{\hbar }{\text{i}}}{\textstyle\frac{\delta }{\delta J}}$
are implied to be substituted instead of the fields, $\Phi$, in $\overline{\Delta }%
M_A$, $M^{B\ast }$, $M_A$, $M^{B_k\ast }_{A}$, $ \overline{\Delta }\Psi_{B}$ in the last equality,
in accordance with the conventions (\ref{MAZ}).

The general result (\ref{varZMf}) for the variation of $Z_{M}$ in the
approximation linear in powers of the variations $\overline{\Delta }\Psi ,%
\overline{\Delta }M$ reads as follows:
\begin{eqnarray}
\overline{\Delta}Z_{M}(J,\Phi^*)& =&  \frac{\im}{\hbar}\Big[\big(J_A+M_{A}(\sfrac{\hbar}{\im}\sfrac{\delta}{\delta
J},\Phi^*)\big) \Big(\frac{\delta}{\delta\Phi^*_A}-\frac{\im}{\hbar}M^{A*}(\sfrac{\hbar}{\im}\sfrac{\delta}{\delta
J},\Phi^*)\Big)\overline{\Delta}\Psi({\textstyle%
\frac{\hbar }{\text{i}}}{\textstyle\frac{\delta }{\delta J}}) \nonumber \\
&&  + \overline{\Delta}M(\sfrac{\hbar}{\im}\sfrac{\delta}{\delta
J},\Phi^*)\Big] Z_{M}(J,\Phi^*),\label{varZMlin}
\end{eqnarray}
and coincides with the result for $\delta Z_{M}$ first obtained in \cite{lrr}.

In the particular case of the absence of BRST symmetry breaking terms, i.e.,
when $M=0$, we derive from (\ref{varZMf}) a new representation for a finite
variation of the functional $Z(J,\Phi ^{\ast })$ under a finite variation of
the gauge condition,
\begin{equation}\label{dZfinBV}
\overline{\Delta }Z(J,\Phi ^{\ast })=\frac{\text{i}}{\hbar
}\left[  \sum_{n\geq 0} \frac{1}{(n+1)!}\Big( \overline{\Delta }\Psi
_{B}({%
\textstyle\frac{\hbar }{\text{i}}}{\textstyle\frac{\delta }{\delta J}})\frac{\delta }{\delta \Phi _{B}^{\ast }}\Big)^n
J_{A}\frac{\delta }{\delta \Phi _{A}^{\ast }}\right]\overline{\Delta }\Psi ({%
\textstyle\frac{\hbar }{\text{i}}}{\textstyle\frac{\delta }{\delta J}}) Z(J,\Phi ^{\ast }),
\end{equation}%
which is reduced, in the case of a small variation, $\overline{\Delta }\Psi
=\delta \Psi $, to the form (\ref{varZBV}), well-known in the BV formalism,
with the notation $\overline{\Delta }Z=\delta Z$.

To complete the research of gauge-dependence in the theory with broken BRST
symmetry, let us calculate the variations of $W_{M}(J,\Phi ^{\ast })$ and $%
\Gamma _{M}(\Phi ,\Phi ^{\ast })$ under a finite change of the gauge
condition, taken into account for the relations $\overline{\Delta }W_{M}={%
\textstyle\frac{\hbar }{\text{i}}}Z_{M}^{-1}\overline{\Delta }Z_{M}$ and $%
\overline{\Delta }W_{M}=\overline{\Delta }\Gamma _{M}$. First of all, $%
\overline{\Delta }W_{M}$ reads%
\begin{eqnarray}
&&\hspace{-1em}\overline{\Delta }W_{M}\ =\ \exp \Big\{\frac{\text{i}}{\hbar }\overline{\Delta }M\big({\textstyle%
\frac{\hbar }{\text{i}}}{\textstyle\frac{\delta }{\delta J}}+{\textstyle\frac{\delta W_M}{\delta J}},\Phi ^{\ast }\big)%
\Big\} \bigg[\overline{\Delta }%
M_A
\sum_{n\geq 1}\frac{1}{n!}\Big\{ \overline{\Delta }\Psi_B
\textstyle\Big(\frac{\delta }{\delta \Phi _{B}^{\ast }}+\frac{\text{i} }{\hbar}\frac{\delta W_M}{\delta \Phi _{B}^{\ast }}-%
\frac{\text{i}}{\hbar }M^{B\ast }\Big)\Big\}^{n-1} \nonumber \\
&& \times \Big(\frac{\delta }{\delta \Phi _{A}^{\ast }}+
\frac{\text{i} }{\hbar}\frac{\delta W_M}{\delta \Phi_{A}^{\ast }}-%
\frac{\text{i}}{\hbar }M^{A\ast }\Big)  - \sum_{n\geq 1}\frac{1}{n!}\Big\{\sum_{k=1}^{n-1}\prod_{l=1}^{k-1}  \overline{\Delta }\Psi_{B_l}\textstyle\Big(\frac{\delta }{\delta \Phi _{B_l}^{\ast }}+\frac{\text{i} }{\hbar}\frac{\delta W_M}{\delta \Phi_{B_l}^{\ast }}-%
\frac{\text{i}}{\hbar }M^{B_l\ast }\Big)\overline{\Delta }\Psi_{B_k}M^{B_k\ast }_{A}  \nonumber\\
&& \times \displaystyle\prod\limits_{l=k+1}^{n-1}  \overline{\Delta }\Psi_{B_l}
\textstyle\Big(\frac{\delta }{\delta \Phi _{B_l}^{\ast }}+\frac{\text{i} }{\hbar}\frac{\delta W_M}{\delta \Phi_{B_l}^{\ast }}-%
\frac{\text{i}}{\hbar }M^{B_l\ast }\Big) \Big\}\textstyle\Big(\frac{\delta }{\delta \Phi _{A}^{\ast }}+\frac{\text{i} }{\hbar}\frac{\delta W_M}{\delta \Phi_{A}^{\ast }}-%
\frac{\text{i}}{\hbar }M^{A\ast }\Big) \nonumber\\
&& + \sum_{n\geq 1}\frac{1}{n!}\Big\{ \overline{\Delta }\Psi_B
\textstyle\Big(\frac{\delta }{\delta \Phi _{B}^{\ast }}+\frac{\text{i} }{\hbar}\frac{\delta W_M}{\delta \Phi_{B}^{\ast }}-%
\frac{\text{i}}{\hbar }M^{B\ast }\Big)\Big\}^{n-1} \Big(J_A+ M_A\Big)\frac{\delta }{\delta \Phi _{A}^{\ast }}
\bigg]\overline{\Delta }\Psi({\textstyle%
\frac{\hbar }{\text{i}}}{\textstyle\frac{\delta }{\delta J}}+{\textstyle\frac{\delta W_M}{\delta J}})
 \nonumber\\
&& +  \sum_{n\geq 1}\frac{1}{n!}\left(\frac{i}{\hbar}\right)^{n-1}\overline{\Delta }M({\textstyle%
\frac{\hbar }{\text{i}}}{\textstyle\frac{\delta }{\delta J}}+{\textstyle\frac{\delta W_M}{\delta J}},\Phi ^{\ast })%
,\label{varWMfin}
\end{eqnarray}%
where use has been made of the Ward identity (\ref{WIWBV}) and substitution of the arguments, $\bigl( {\textstyle%
\frac{\hbar }{\text{i}}}{\textstyle\frac{\delta }{\delta J}}+{\textstyle\frac{\delta W_M}{\delta J}}\bigr)$, instead of $\Phi$ in $\overline{\Delta }%
M_A$, $M^{B\ast }$, $M_A$, $ \overline{\Delta }\Psi_{B}$ should be made. Again, without
BRST symmetry breaking terms for $M=0$, we can obtain from Eq. (\ref{varWMfin})
a new representation for a finite variation for the generating functional $%
W(J,\Phi ^{\ast })$, namely,%
\begin{equation}\label{dWfinBV}
\overline{\Delta }W \ =\ \left[  \sum_{n\geq 0} \frac{1}{(n+1)!}\Big( \overline{\Delta }\Psi
_{B}\big({\textstyle\frac{\delta W}{%
\delta J}}+{\textstyle\frac{\hbar }{\text{i}}}{\textstyle\frac{\delta }{%
\delta J}}\big)\Big[\frac{ \text{i}}{\hbar}\frac{\delta W}{\delta \Phi _{B}^{\ast }}+\frac{\delta }{\delta \Phi _{B}^{\ast }}\Big]\Big)^n
J_{A}\frac{\delta }{\delta \Phi _{A}^{\ast }}\right]\overline{\Delta }\Psi  \big({\textstyle\frac{\delta W}{%
\delta J}}+{\textstyle\frac{\hbar }{\text{i}}}{\textstyle\frac{\delta }{%
\delta J}}\big).
\end{equation}%
For the first order in powers of $\overline{\Delta }\Psi $ and $\overline{%
\Delta }M$, the variation (\ref{varWMfin}) for $\overline{\Delta }%
W_{M}(J,\Phi ^{\ast })$ has the form
\beq
\label{varWlin} \overline{\Delta} W_M  =
\Big(J_A+M_{A}\big(\sfrac{\delta W_M}{\delta
J}+\sfrac{\hbar}{\im}\sfrac{\delta}{\delta
J},\Phi^*\big)\Big)\frac{\delta }{\delta\Phi^*_A} \overline{\Delta}\Psi
\big(\sfrac{\delta W_M}{\delta
J}+\sfrac{\hbar}{\im}\sfrac{\delta}{\delta J}\big) \ +\ \overline{\Delta}
M\big(\sfrac{\delta W_M}{\delta
J}+\sfrac{\hbar}{\im}\sfrac{\delta}{\delta J},\Phi^*\big) \,,
\eeq
identical (after the change $\overline{\Delta }\rightarrow \delta $) with
the one obtained in \cite{llr}, \cite{lrr}.

Second, on order to derive a finite form of the gauge variation for the
effective action, we can use the calculations of \cite{lrr}. Namely, the
change of variables $(J_{A},\Phi _{A}^{\ast })\rightarrow (\Phi ^{A},\Phi
_{A}^{\ast })$ from the Legendre transformation (\ref{EAq}) implies
\beq
\label{dphistar}
{\frac{\delta}{\delta\Phi^*}}\Big|_{J}=
\frac{\delta}{\delta\Phi^*}\Big|_{\Phi} + \frac{\delta
\Phi}{\delta\Phi^*}{\frac{\delta_{\it
l}}{\delta\Phi}}\Big|_{\Phi^*}\texttt{ and }\frac{\delta W_M}{\delta\Phi^*_A} \ =\ \frac{\delta \Gamma_M}{\delta\Phi^*_A}.
\eeq
Then, the differential consequence of the Ward identities for $Z_{M}$~(\ref%
{wardZM}) and $W_{M}$~(\ref{WIWBV}) implies%
\begin{eqnarray}
-\Big(\frac{\delta \Gamma _{M}}{\delta \Phi ^{A}}-{\widehat{M}}_{A}\Big)%
\frac{\delta \Phi ^{B}}{\delta \Phi _{A}^{\ast }} &=&-\Big(\frac{\delta
\Gamma }{\delta \Phi _{B}^{\ast }}-{\widehat{M}}^{B^{\ast }}\Big)%
(-1)^{\varepsilon _{B}}  \notag  \label{dGamma} \\
&&+\frac{i}{\hbar }\Big[-{\widehat{M}}_{A}\frac{\delta \Gamma _{M}}{\delta
\Phi _{A}^{\ast }}-\frac{\delta \Gamma _{M}}{\delta \Phi ^{A}}{\widehat{M}}%
^{A^{\ast }}+{\widehat{M}}_{A}{\widehat{M}}^{A^{\ast }},\Phi ^{B}\Big],
\end{eqnarray}%
with the same notation $\big[\ ,\ \big]$ for the supercommutator
as in (\ref{spsi+dp}).

Using Eqs. (\ref{varWMfin}), (\ref{dphistar}), (\ref{dGamma}) and the
relation
\beq
\label{phiphi*} \frac{\delta \Phi^B}{\delta\Phi^*_A} =
(-1)^{\vp_B(\vp_A+1)}\frac{\delta }{\delta J_B} \frac{\delta W}{
\delta \Phi^*_A} =- (-1)^{\vp_B(\vp_A+1)}(\Gamma^{''-1})^{BC}
\frac{\delta_{\it l} }{\delta \Phi^C}\frac{\delta \Gamma}{ \delta
\Phi^*_A},
\eeq
we present the finite variation of the effective action in the form
\begin{eqnarray}
\overline{\Delta }\Gamma _{M} &=&
\exp \Big\{\frac{\text{i}}{\hbar }\langle\overline{\Delta }M\rangle%
\Big\} \bigg(\langle\overline{\Delta }%
M_A\rangle
\sum_{n\geq 1}\frac{1}{n!}\Big\{ \langle\overline{\Delta }\Psi_B\rangle
\textstyle\Big(- {\widehat F}^B +\frac{\text{i} }{\hbar}\frac{\delta \Gamma_M}{\delta \Phi _{B}^{\ast }}-  %
\frac{\text{i}}{\hbar }\widehat{M}{}^{B\ast }\Big)\Big\}^{n-1} \nonumber \\
&& \times \textstyle \Big(- {\widehat F}^A +\frac{\text{i} }{\hbar}\frac{\delta \Gamma_M}{\delta \Phi _{A}^{\ast }}-  %
\frac{\text{i}}{\hbar }\widehat{M}{}^{A\ast }\Big)  -  \displaystyle\sum_{n\geq 1}\frac{1}{n!}\Big\{\sum_{k=1}^{n-1}\prod_{l=1}^{k-1}  \langle\overline{\Delta }\Psi_{B_l}\rangle\textstyle\Big(- {\widehat F}^{B_l} +\frac{\text{i} }{\hbar}\frac{\delta \Gamma_M}{\delta \Phi_{B_l}^{\ast }}-  %
\frac{\text{i}}{\hbar }\widehat{M}{}^{B_l\ast }\Big) \nonumber\\
&& \times\langle\overline{\Delta }\Psi_{B_k}\rangle \widehat{M}{}^{B_k\ast }_{A} \prod\limits_{l=k+1}^{n-1}  \langle\overline{\Delta }\Psi_{B_l}\rangle
\textstyle\Big(- {\widehat F}^{B_l} +\frac{\text{i} }{\hbar}\frac{\delta \Gamma_M}{\delta \Phi _{B_l}^{\ast }}-  %
\frac{\text{i}}{\hbar }\widehat{M}{}^{B_l\ast }\Big) \Big\}\textstyle\Big(- {\widehat F}^A +\frac{\text{i} }{\hbar}\frac{\delta \Gamma_M}{\delta \Phi _{A}^{\ast }}-  %
\frac{\text{i}}{\hbar }\widehat{M}{}^{A\ast }\Big) \nonumber\\
&& + \sum_{n\geq 1}\frac{1}{n!}\Big\{ \langle\overline{\Delta }\Psi_B\rangle
\textstyle\Big(- {\widehat F}^B +\frac{\text{i} }{\hbar}\frac{\delta \Gamma_M}{\delta \Phi _{B}^{\ast }}-  %
\frac{\text{i}}{\hbar }\widehat{M}{}^{B\ast }\Big)\Big\}^{n-1}  \Big\{-\big(\Gamma _{M},\ \ \big)\ +\ \displaystyle{\widehat{M}}_{A}\frac{\delta }{\delta \Phi _{A}^{\ast }}  \notag \\
&& +\ (-1)^{\varepsilon _{A}}{\widehat{M}}^{A\ast
}\frac{\delta _{l}}{\delta \Phi ^{A}} -\ \frac{\text{i}}{\hbar }\Big[{\widehat{M}}_{A}\frac{\delta \Gamma _{M}}{%
\delta \Phi _{A}^{\ast }}+\frac{\delta \Gamma _{M}}{\delta \Phi ^{A}}{%
\widehat{M}}^{A\ast }-{\widehat{M}}_{A}{\widehat{M}}^{A\ast },\ \Phi ^{B}%
\Big]\frac{\delta _{\mathit{l}}}{\delta \Phi ^{B}}\, \Big\}
\bigg)\langle \overline{%
\Delta }\Psi \rangle
 \nonumber\\
&& +  \sum_{n\geq 1}\frac{1}{n!}\left(\frac{i}{\hbar}\right)^{n-1}\langle\overline{\Delta }M\rangle
 .
\label{varGammaM}
\end{eqnarray}%
Here, we use the notation
\beq \label{laradepsi}\langle\overline{\Delta}\Psi\rangle\=\overline{\Delta}\Psi({\widehat\Phi})\cdot 1
\und \langle\overline{\Delta} M\rangle\=\overline{\Delta} M({\widehat \Phi},\Phi^*)\cdot
1 \ ,\eeq
as well as  the same notation
for $\langle\overline{\Delta }\Psi_{B_k}\rangle$, $\langle\overline{\Delta }%
M_A\rangle$, and   introduce the operator ${\widehat{F}}^{A}$, derived from Eqs. (\ref{dphistar}), (%
\ref{dGamma}), (\ref{phiphi*}), as follows%
\beq {\widehat F}^A &=&-\frac{\delta}{\delta\Phi^*_A}\ +\
(-1)^{\vp_B(\vp_A+1)} (\Gamma_M^{''-1})^{BC}\Big(\frac{\delta_{\it
l}}{\delta\Phi^C}\frac {\delta
\Gamma_M}{\delta\Phi^{*}_{A}}\Big)\frac{\delta_{\it l}
}{\delta\Phi^B}\ . \label{FAdef} \eeq

Now, we can deduce from Eq. (\ref{varGammaM}) a new representation for a finite
variation of the effective action $\Gamma (J,\Phi ^{\ast })$ in a local form
without BRST symmetry breaking terms ($M=0$),%
\begin{eqnarray}
\overline{\Delta }\Gamma & = & - \sum_{n\geq 0} \frac{1}{(n+1)!}\Big( \langle\overline{\Delta }\Psi
_{B}\rangle \Big[\frac{ \text{i}}{\hbar}\frac{\delta \Gamma}{\delta \Phi _{B}^{\ast }}- {\widehat F}^B\Big]\big|_{M=0}\Big)^n
\big(\Gamma ,\langle \overline{\Delta }\Psi \rangle \big), \label{vGammafinBV}
\end{eqnarray}%
in the first order with respect to the variation $\langle \overline{\Delta }%
\Psi \rangle $, identical with the previously known representation (\ref{varGBV}).

For the first order in powers of $\langle \overline{\Delta }\Psi \rangle $
and $\langle \overline{\Delta }M\rangle $, the variation (\ref{varGammaM})
of $\overline{\Delta }\Gamma _{M}(J,\Phi ^{\ast })$ takes the previously
known \cite{lrr} \textquotedblleft local-like\textquotedblright\ form
\begin{eqnarray}
\overline{\Delta }\Gamma _{M} &=&-\big(\Gamma _{M},\langle \overline{\Delta }%
\Psi \rangle \big)\ +\ \left( {\widehat{M}}_{A}\frac{\delta }{\delta \Phi _{A}^{\ast }}+\
(-1)^{\varepsilon _{A}}{\widehat{M}}^{A\ast }\frac{\delta _{l}}{\delta \Phi
^{A}}\right) \langle \overline{\Delta }\Psi \rangle  \notag \\
&&-\ \frac{\text{i}}{\hbar }\Big[{\widehat{M}}_{A}\frac{\delta \Gamma _{M}}{%
\delta \Phi _{A}^{\ast }}+\frac{\delta \Gamma _{M}}{\delta \Phi ^{A}}{%
\widehat{M}}^{A\ast }-{\widehat{M}}_{A}{\widehat{M}}^{A\ast },\ \Phi ^{B}%
\Big]\frac{\delta _{\mathit{l}}}{\delta \Phi ^{B}}\,\langle \overline{%
\Delta }\Psi \rangle + \langle \overline{%
\Delta }M\rangle \ ,  \label{varGMlin}
\end{eqnarray}%
where coincidence with the final result of \cite{lrr} is achieved by the
change $\overline{\Delta }\rightarrow \delta $.

To study gauged (field-dependent) BRST transformations in a theory with broken BRST
symmetry, we follow the result of \cite{llr}, \cite{lrr} and present the
variation, linear in $\langle \overline{\Delta }\Psi \rangle $, $\langle
\overline{\Delta }M\rangle $, of the effective action (\ref{varGMlin}) in an
equivalent form, being the so-called non-local form, due to the explicit presence
of $(\Gamma _{M}^{^{\prime \prime }-1})^{BC}$ in ${\widehat F}^A$ (\ref{FAdef}):
\beq
\overline{\Delta}\Gamma_M \= \frac{\delta\Gamma_M}{\delta\Phi^A}
{\widehat F}^A\,\langle\overline{\Delta}\Psi\rangle\ -\ {\widehat
M}_A{\widehat F}^A \langle\overline{\Delta}\Psi\rangle\ +\ \langle\overline{\Delta}
M\rangle\ . \label{varGammaFlin}
\eeq

We now intend to revise our previous result \cite{llr}, \cite{lrr}, which states
that the variation (\ref{varGammaFlin}) implies that the effective action
with soft BRST symmetry breaking is generally gauge-dependent on the mass shell,
since
\begin{equation}\label{old}
\frac{{\delta }\Gamma _{M}}{\delta \Phi ^{A}}=0\qquad \longrightarrow \qquad
\overline{\Delta }\Gamma _{M}\neq 0\ .
\end{equation}%
Indeed, there is a hope that the introduction of broken BRST symmetry into
the field-antifield formalism would be consistent only if the two final
terms in (\ref{varGammaFlin}) should cancel each other:%
\beq \label{BasRest}
\langle\overline{\Delta} M\rangle\={\widehat M}_A{\widehat F}^A \langle\overline{\Delta}\Psi\rangle\ ,
\eeq
which, at the classical level, imposes a condition on the gauge variation of $%
M $ under a change of the gauge-fixing functional~$\Psi $,
\begin{equation}\label{classrestM}
\overline{\Delta }M \ =\ {\frac{\delta M}{\delta \Phi ^{A}}}\,{\widehat{F}}%
_{0}^{A}\,\overline{\Delta }\Psi \quad \mathrm{where}\quad {\widehat{F}}%
_{0}^{A}\ =\ {(}-1)^{\varepsilon _{B}(\varepsilon _{A}+1)}(S^{^{\prime \prime
}-1})^{BC}\Big(\frac{\delta _{\mathit{l}}}{\delta \Phi ^{C}}\frac{\delta S}{%
\delta \Phi _{A}^{\ast }}\Big)\frac{\delta _{\mathit{l}}}{\delta \Phi ^{B}}\
.
\end{equation}%
Of course, despite the fact that it seems to be a strong restriction that the
BRST-breaking functional~$M$ corresponding to the effective action should be
gauge-independent on the mass shell (implying the gauge-independence
of the physical S-matrix), the gauge-independence (but not invariance)
can, in fact, be restored.

In order to justify the above proposition, let us subject the integrand in
$Z_{M}$, with the gauge-fixing functional $\Psi (\Phi )$, to the change of
variables (\ref{BRSTgen}), with a field-dependent odd-valued parameter
$\Lambda\left(\Phi,\Phi ^{\ast }|\overline{\Delta} \Psi\right)$ in
(\ref{solcompeq}) being a solution of Eq.~(\ref{eqlambdapsi})
[corresponding to the functional
$\hat{\Lambda}(\Phi ^{\prime \prime })$ from Eq. (\ref{aollambda})],
which  provides the gauge-independence
of the vacuum functional $Z_{\Psi}(0,\Phi ^{\ast })$  (\ref{equival}):%
\begin{eqnarray}
Z_{M,\Psi}(J,\Phi ^{\ast }) &=&\int \!D\Phi \ \exp \Big\{\frac{\text{i}}{\hbar }%
\Big( S -i\hbar \,\mbox{ln}\,\Big[\big(1+s_{e}\Lambda\left(\overline{\Delta} \Psi\right) \big)^{-1} \big(1+\overleftarrow{s}_{e}\Lambda\left(\overline{\Delta} \Psi\right) \big) \Big]   \notag
 \\
&& + s_{e}M(\Phi ,\Phi ^{\ast })\Lambda\left(\overline{\Delta} \Psi\right)
 + J_{A}\big[\Phi ^{A}+(s_{e}\Phi ^{A})\Lambda\left(\overline{\Delta} \Psi\right)\big]\Big)%
\Big\}.\label{changZ1}
\end{eqnarray}%
Thus, taking into account the fact that for any variation of the gauge-fixing
functional, $\Psi (\Phi )\rightarrow (\Psi +\overline{\Delta }\Psi )(\Phi )$%
, in view of the result obtained in Section~\ref{gdepBRSTtrans}, there
exists a parameter, $\Lambda\left(\overline{\Delta} \Psi\right)$, of finite gauged BRST
transformations, being a solution of Eq. (\ref{eqlambdapsi}) such that the
 action $S_{\Psi }$, and therefore also the total quantum action $S$
in the gauge determined by $\Psi +\overline{\Delta }\Psi $, takes the form
\begin{eqnarray}
&&\hspace{-1em}S_{\Psi +\overline{\Delta }\Psi }+M_{\Psi +\overline{\Delta }%
\Psi }\ =\ \Big[S_{\Psi }- i\hbar \,\mbox{ln}\,\Big\{\big(1+s_{e}\Lambda\left(\overline{\Delta} \Psi\right) \big)^{-1} \big(1+\overleftarrow{s}_{e}\Lambda\left(\overline{\Delta} \Psi\right) \big) \Big\}\Big]
\notag  \label{totadelpsi} \\
&&+\left[M_\Psi+s_{e}M_\Psi\left(\Phi ,\Phi ^{\ast
}\right)\Lambda\left(\overline{\Delta} \Psi\right)\right],
\end{eqnarray}%
where the first square brackets in the right-hand-side contain an expression for $%
S_{\Psi +\overline{\Delta }\Psi }$, whereas the second brackets
should contain an expression for $M_{\Psi +\overline{\Delta }} = M_{\Psi }+ \overline{\Delta }M$. Therefore, based on the equivalence
theorem \cite{FradkinTyutin}, we have, first, a representation, being different
form (\ref{varZMf}), for a finite change of the functional
$Z_{M,\Psi}(J,\Phi ^{\ast })$ under a finite change of the gauge:
\begin{eqnarray}
\overline{\Delta} Z_{M,\Psi}(J,\Phi ^{\ast }) &=& - \frac{\text{i}}{\hbar }J_{A}\big(s_{e}\Phi ^{A}\big)\left({%
\textstyle\frac{\hbar }{\text{i}}}{\textstyle\frac{\delta }{\delta J}},\Phi^*\right)\Lambda\left({%
\textstyle\frac{\hbar }{\text{i}}}{\textstyle\frac{\delta }{\delta J}},\Phi^*|\overline{\Delta} \Psi\right)  Z_{M,\Psi}(J,\Phi ^{\ast }) \nonumber \\
&=& (-1)^{\varepsilon_A} J_{A}\Lambda\left({%
\textstyle\frac{\hbar }{\text{i}}}{\textstyle\frac{\delta }{\delta J}},\Phi^*|\overline{\Delta} \Psi\right)\left(\frac{\delta}{\delta\Phi^*_A} -\frac{\text{i}}{\hbar } M^{A*}_\Psi \right)  Z_{M,\Psi}(J,\Phi ^{\ast })\label{finchangZ1}
\end{eqnarray}%
which also leads to the on-shell coincidence (for $J_A=0$) of the
generating functionals $Z_{M}(J,\Phi ^{\ast })$ calculated in the gauges $%
\Psi $ and $\Psi +\overline{\Delta }\Psi $, respectively; we also obtain
the form of the soft BRST symmetry functional $M$ in the gauge determined
by ${\Psi +\overline{\Delta }\Psi }$, provided that in the gauge $\Psi $
the former is defined by the functional $M$,
\beq\label{Mdeltalin}
 \overline{\Delta} M \ =\  (s_e M)\Lambda\left(\overline{\Delta} \Psi\right).
\eeq
In the approximation linear in $\overline{\Delta }\Psi $, we have, making use of
(\ref{Mdeltalin}) and (\ref{eqlambdapsilin}),
\beq\label{Mdelta}
 M_{\Psi+\overline{\Delta}\Psi} \ =\  M  - \frac{i}{\hbar} M_{A}S^A_\Psi
{\delta}\Psi.
\eeq
An important particular case, which covers practically all the known gauge
models, corresponds to a gauge theory of first rank with a closed
algebra, when Eqs. (\ref{conseqhatS}) for the quantum action are fulfilled.
An explicit expression of the soft BRST-breaking functional similar to Eq.
(\ref{Mdelta}) in the gauge $(\Psi +\overline{\Delta }\Psi )$ reads as
follows:
\beq\label{Mdelta1r}
 M_{\Psi+\overline{\Delta}\Psi} \ =\  M + M_{A}S^A_\Psi \overline{\Delta}\Psi \big(s\overline{\Delta}\Psi\big)^{-1}\Big[\exp \left\{-\frac{i}{\hbar} s\big(\overline{\Delta}\Psi\big)\right\} -1\Big],
\eeq
where account has been taken of Eq. (\ref{aollambda}).

Note, first of all, that the additional contribution to $M$ in (\ref%
{Mdelta1r}) does not increase the maximal power in the antifields of the
functional $M$. Second, the gauge variation of the BRST-symmetry-broken
functional does not generally turn the solution of the soft BRST-breaking
equation (\ref{brbrst}) into a solution. However, in the above case of a
gauge theory with closed algebra (reducible or not) of first rank with $%
M=M(\Phi )$, Eq. (\ref{brbrst}) for the gauge-transformed functional $%
M_{\Psi +\overline{\Delta }\Psi }$ is valid by construction, due to the
representation (\ref{Mdelta1r}).

Let us now check the validity of the representation (\ref{BasRest}) for a
variation of the BRST symmetry breaking functional with accuracy up to the
first order in the gauge variation $\overline{\Delta }\Psi $.
In fact, it is sufficient to compare two representations for a finite change of $Z_{M,\Psi}(J,\Phi ^{\ast })$,
with (\ref{varZMlin}) obtained  from a change of the gauge condition and with
(\ref{finchangZ1})
obtained via a change  of variables generated by field-dependent
 BRST transformations with the parameter $\Lambda\left(\overline{\Delta} \Psi\right)$.
Indeed, (\ref{varZMlin}) can be presented as
\begin{eqnarray}
\overline{\Delta}Z_{M,\Psi}& =&  \frac{\im}{\hbar} J_A \Big(\frac{\delta}{\delta\Phi^*_A}-\frac{\im}{\hbar}M^{A*}(\sfrac{\hbar}{\im}\sfrac{\delta}{\delta
J},\Phi^*)\Big) \overline{\Delta}\Psi({\textstyle%
\frac{\hbar }{\text{i}}}{\textstyle\frac{\delta }{\delta J}})Z_{M,\Psi}\label{varZMlinprov}   \\
&+& \frac{\im}{\hbar}\Big[M_{A}(\sfrac{\hbar}{\im}\sfrac{\delta}{\delta
J},\Phi^*)\Big(\frac{\delta}{\delta\Phi^*_A}-\frac{\im}{\hbar}M^{A*}(\sfrac{\hbar}{\im}\sfrac{\delta}{\delta
J},\Phi^*)\Big)\overline{\Delta}\Psi({\textstyle%
\frac{\hbar }{\text{i}}}{\textstyle\frac{\delta }{\delta J}})  + \overline{\Delta}M(\sfrac{\hbar}{\im}\sfrac{\delta}{\delta
J},\Phi^*)\Big] Z_{M,\Psi},\nonumber
\end{eqnarray}
and for the summand in the second line we have
\begin{eqnarray}
&& \frac{\im}{\hbar}\Big[\frac{\im}{\hbar}M_{A}(\sfrac{\hbar}{\im}\sfrac{\delta}{\delta
J},\Phi^*)S_\Psi^{A}(\sfrac{\hbar}{\im}\sfrac{\delta}{\delta
J},\Phi^*)\overline{\Delta}\Psi({\textstyle%
\frac{\hbar }{\text{i}}}{\textstyle\frac{\delta }{\delta J}})  + \overline{\Delta}M(\sfrac{\hbar}{\im}\sfrac{\delta}{\delta
J},\Phi^*)\Big] Z_{M,\Psi}\nonumber\\
&& = \  \frac{\im}{\hbar}\Big[\frac{\im}{\hbar}(s_e M)(\sfrac{\hbar}{\im}\sfrac{\delta}{\delta
J},\Phi^*) \overline{\Delta}\Psi({\textstyle%
\frac{\hbar }{\text{i}}}{\textstyle\frac{\delta }{\delta J}})+ \overline{\Delta}M(\sfrac{\hbar}{\im}\sfrac{\delta}{\delta
J},\Phi^*)\Big] Z_{M,\Psi} \nonumber\\
&& = \  \frac{\im}{\hbar}\Big[\overline{\Delta}M - (s_e M)\Lambda\left(%
\overline{\Delta}\Psi\right)\Big] (\sfrac{\hbar}{\im}\sfrac{\delta}{\delta
J},\Phi^*) Z_{M,\Psi} = 0    \label{varZMlinprov1}
\end{eqnarray}
due to the representation (\ref{Mdeltalin})
for $\overline{\Delta}M $.
The last expression in terms of the average fields (\ref{EAq})
for the effective action $\Gamma_M$ is nothing else
than the representation (\ref{BasRest}).

Thus, the coincidence
of  $\overline{\Delta}Z_{M,\Psi}(J,\Phi^*)$ in (\ref{varZMlin})
with (\ref{finchangZ1}) is guaranteed due to (\ref{Mdeltalin}) and (\ref{solcompeq}).

As a consequence, the finite change of the functionals $W_{M}$, $\Gamma _{M}$ in
the linear approximation in $\overline{\Delta}\Psi$ in the relations (\ref{varWlin})
and (\ref{varGammaFlin}) should coincide
(after change $\overline{\Delta }\rightarrow \delta $)
with  the variations of the functionals $W$, $\Gamma$, respectively,
in (\ref{varWBV}) and (\ref{varGammaBV})
without any soft BRST symmetry breaking term $M$.
Concerning the finite change of the $W_{M}$, $\Gamma _{M}$ in
(\ref{varWMfin}) and (\ref{varGammaM}),
as a result of the above-established correspondence between
the finite change of the gauge
$\overline{\Delta}\Psi$
and the parameter of gauged BRST transformation $\Lambda\left(\Phi,\Phi^*|%
\overline{\Delta}\Psi\right)$ in (\ref{solcompeq}),
the form of $\overline{\Delta}M$ should be chosen according to (\ref{Mdeltalin}).

Thus, we have proved the following \textbf{\emph{Statement}}: an addition to the
quantum action $S_{\Psi }$, satisfying the master equation in the BV
formalism (\ref{MEBV}), of a  term, $M(\Phi ,\Phi ^{\ast })$, breaking the BRST
symmetry softly\footnote{Of course,
any such term should be admissible,
in order to have a well-determined
path integral, at least in
perturbation theory. Second,
the requirement of soft breaking
of the  BRST  symmetry may be weakened
in order to consider only the breaking
of BRST symmetry
(see Footnote~6 for details).}
(\ref{brbrst}), first, destroys the BRST invariance
of the integrand in the generating functional of Green's functions, $Z_{M}$%
, and therefore also the gauge-invariance of the total action $(S_{\Psi
}+M)$ in the tree approximation; second, this leads to an
effective action, $\Gamma _{M}$, being gauge-independent upon a
variation of the gauge condition within the class of admissible
gauges on its extremals,
\beq\label{finvarGM} \frac{\delta\Gamma_M}{\delta\Phi^A}=0\texttt{ and } \langle\overline{\Delta}M\rangle  ={\widehat M}_A{\widehat F}^A \langle\overline{\Delta}\Psi\rangle
\longrightarrow\qquad \overline{\Delta}\Gamma_M = 0\ ,
\eeq
providing the variation of the BRST symmetry breaking term $M$ in the form
of gauged (field-dependent) BRST symmetry transformations (\ref{Mdeltalin}).

In particular, this implies that if in
the reference frame determined by the gauge fermion $\Psi$
the generating functional $Z_{M,\Psi}(J,\Phi^*)$
is described by (\ref{ZSfull})
then in the reference frame  $\Psi+\overline{\Delta}\Psi$ it should have the form
\beq \label{ZSfullch}
Z_{M+\overline{\Delta}, \Psi+\overline{\Delta}\Psi}(J,\Phi^*)\ = \int\!D\Phi\ \exp \Big\{\frac{\im}{\hbar}
\Big(S_{\Psi+\overline{\Delta}\Psi}+ M_\Psi + \big(s_e M_\Psi\big)\Lambda\left(%
\overline{\Delta}\Psi\right) + J_A\Phi^A\Big)\Big\}\ . \eeq
This fact makes the procedure of Lagrangian quantization of a gauge theory
with soft BRST symmetry breaking
consistent\footnote{However, another basic requirement
for a quantum gauge field theory, i.e., the unitarity of the
S-matrix, is destroyed when adding to the gauge theory
any soft BRST symmetry breaking terms, and thus needs
a special investigation.} and leads, in particular, to a
gauge-independent S-matrix within the conventional
approach \cite{books, Bogolyubov, FaddeevSlavnov}.
The problem of gauge dependence considered in a non-renormalized
gauge theory with BRST broken terms should now be studied
in a renormalized theory.

Let us turn to some field-theoretic examples and constructions where
the concept of BRST symmetry breaking is realized.

%%%%%%%%%%%%%%%%%%%%%%%%%%%%%%%%%%%%%%%%%%%%%%%%%%%

\section{Application to Functional Renormalization Group}

\label{FRGandPV}

\noindent

In this section, we apply the above results to the study of the effective
average action proposed in \cite{Wett-1}, \cite{Wett-Reu-1}, \cite{Wett-Reu-2}, which naturally arises within the functional renormalization
group (FRG) approach to the Lagrangian quantization of Yang--Mills theories
and was recently examined in \cite{LS}.

The essence of FRG is to use, instead of $\Gamma $, the so-called
effective average action $\Gamma _{k}$ with a momentum-shell parameter, $k$%
, coinciding with $\Gamma $ for vanishing $k$,
\begin{equation}\label{limgk}
\lim_{k\rightarrow 0}\Gamma _{k}=\Gamma ,
\end{equation}%
in such a way that the Faddeev--Popov action $S_{FP}(\Phi )$ for
Yang--Mills theories should be extended by means of soft BRST
symmetry breaking terms, $M$, having the form of the regulator
action $S_{k}$ for $M=S_{k}$,
\begin{eqnarray}
S_{k}(A,C,{\bar{C}}) &=&\frac{1}{2}A^{i}A^{j}(R_{k,A})_{ij}+{\bar{C}}%
^{\alpha }(R_{k,gh})_{\alpha \beta }C^{\beta }  \notag  \label{regSk} \\
&=&\int d^{D}x\Big\{\frac{1}{2}A^{a\mu }(x)(R_{k,A})_{\mu \nu
}^{ab}(x)A^{b\nu }(x)+{\bar{C}}^{a}(x)(R_{k,gh})^{ab}(x)C^{b}(x)\Big\}.
\end{eqnarray}%
In (\ref{regSk}), we have specified the condensed notations, so
that the total configuration space $\mathcal{M}$ of the
Yang--Mills theory,
\begin{equation*}\label{confYM}
\{\Phi ^{A}\}={\{}A^{i},C^{\alpha },{\bar{C}}^{\alpha },B^{\alpha
}\}=\{A_{\mu }^{a},C^{a},{\bar{C}}^{a},B^{a}\}(x)\quad \varepsilon
(C^{a})=\varepsilon (\bar{C}^{a})=1\ ,\ \varepsilon (A_{\mu
}^{a})=\varepsilon (B^{a})=0\ ,
\end{equation*}%
which determines this irreducible theory and includes the fields $A_{\mu }^{a}(x)$,
the Grassmann-odd Faddeev--Popov (scalar) ghost and antighost fields
$C^{a}$ and ${\bar{C}}^{a}$, as well as the Nakanishi--Lautrup auxiliary
fields $B^{a}$, given in the $D$-dimensional Minkowski space-time $%
R^{1,D-1}$ and taking values in the adjoint representation of the Lie algebra~$%
su(N)$. In turn, the regulator quantities $(R_{k,A})$,
$(R_{k,gh})$, having no dependence on the fields, obey the property
$(R_{k,A})_{ij}=(-1)^{\varepsilon _{i}\varepsilon
_{j}}(R_{k,A})_{ji}$ and vanish as the parameter $k$ tends to zero.

The initial classical action $S_{0}$ of the Yang--Mills fields $A_{\mu
}^{a}(x)$ and its gauge transformations have the standard form
(with the coupling constant $g=1$, for simplicity)
\begin{eqnarray}  \label{gtrYM}
&&S_{0}(A) = {-}\frac{1}{4}\int d^{D}x\ F_{\mu \nu }^{a}F^{\mu \nu {}a}\ %
\mathrm{ for }\quad F_{\mu \nu }^{a} = {\partial}_{\mu }A_{\nu
}^{a}-\partial _{\nu }A_{\mu }^{a}+f^{abc}A_{\mu }^{b}A_{\nu }^{c}\ ,
\label{clYM} \\
&&\quad \delta A_{\mu }^{a}  = {D}_{\mu }^{ab}\xi ^{b},\qquad D_{\mu }^{ab}%
 = {\delta}^{ab}\partial _{\mu }+f^{acb}A_{\mu }^{c},\quad \varepsilon (\xi
^{b})=0,
\end{eqnarray}%
with the Lorentz indices $\mu ,\nu =0,1,...,D{-}1$, the metric
tensor $\eta _{\mu \nu }$, $\eta _{\mu \nu
}=\mathrm{diag}(-,+,...,+)$, the totally antisymmetric~$su(N)$
structure constants $f^{abc}$, for $a,b,c = 1,\ldots, N^2-1$, the covariant derivative $D_{\mu
}^{ab}$, and arbitrary functions $\xi^{b}$ in $R^{1,D-1}$.

The corresponding set of odd momenta for the fields, i.e., antifields, reads%
\begin{equation*}
\{\Phi _{A}^{\ast }\} = {\{}A^{\ast a\mu },C^{\ast a},{\bar{C}}^{\ast
a},B^{\ast a}\}(x)\qquad \mathrm{with}\quad \varepsilon (A^{\ast a\mu
})=\varepsilon (B^{\ast a})=1\ ,\quad \varepsilon (C^{\ast a})=\varepsilon ({%
\bar{C}}^{\ast a})=0\ ,
\end{equation*}%
whereas a solution $\bar{S}(\Phi ,\Phi ^{\ast })$ to the quantum
master equation (\ref{MEBV}) can be presented as%
\begin{equation*}
\bar{S}(\Phi ,\Phi ^{\ast })= {S}_{0}(A)\ +\ \int d^{D}x\
\Big\{A^{\ast a\mu }D_{\mu }^{ab}C^{b}\ +\
{\textstyle\frac{1}{2}}C^{\ast a}f^{abc}C^{b}C^{c}\ +\
\bar{C}{}^{\ast a}B^{a}\Big\}\ ,
\end{equation*}%
which, in view of the identity $\Delta \bar{S}=0$, is also a solution to the classical
master equation $(\bar{S},\bar{S})=0$. The gauge-fixed action $%
S_{\Psi }(\Phi ,\Phi ^{\ast })=\bar{S}(\Phi ,\Phi ^{\ast }+{\textstyle\frac{%
\delta \Psi }{\delta \Phi }})$ obeys the same equations with a Grassmann-odd
gauge-fixing functional $\Psi (\Phi )$, which can be chosen as
\beq
\Psi(\Phi)\={\bar C}^a\chi^a(A,B)\ \ \mathrm{ with }\ \ \chi^a \= 0
\label{psiGk}\eeq
so that the non-renormalized Faddeev--Popov action $S_{FP}(\Phi )$ is
obtained from $S_{\Psi }$ for vanishing antifields, $\Phi _{A}^{\ast }$,
\begin{eqnarray}
S_{FP}(\Phi ) &=&\Big[1-\Phi _{A}^{\ast }\frac{\delta }{\delta \Phi
_{A}^{\ast }}\Big]S_{\Psi }(\Phi ,\Phi ^{\ast })\ =S_{0}(A)\ +\ {\bar{C}}%
^{a}K^{ab}C^{b}\ +\ \chi ^{a}B^{a}  \notag \\
&=&S_{0}(A)\ +\ {s}\Psi (\Phi ),  \label{SFPym}
\end{eqnarray}%
where $K^{ab}$ and ${s}$ are the Faddeev--Popov operator and the
Slavnov variation (\ref{Slavnovvarr}), written for any functional
$F(\Phi )$ as follows:
\begin{equation}\label{FPoper}
K^{ab}={\frac{\delta \chi ^{a}}{\delta A_{\mu }^{c}}}D_{\mu }^{cb}\ \ \
\mathrm{and\ \ \ }sF(\Phi )\ =\ \frac{\delta F}{\delta \Phi ^{A}}\frac{%
\delta S_{\Psi }}{\delta \Phi _{A}^{\ast }}\ .
\end{equation}%

Both actions $S_{\Psi }(\Phi ,\Phi ^{\ast })$, $S_{FP}(\Phi )$ are
invariant with respect to the BRST transformation [compare with
Eq. (\ref{BRSTc})]
\begin{equation*}
\delta _{\mu }\Phi ^{A}=S_{\Psi }^{A}\mu \ \ \ \mathrm{with\ \ \ }S_{\Psi
}^{A}\ =\big(D_{\mu }^{ab}C^{b},\ {\textstyle\frac{1}{2}}f^{abc}C^{b}C^{c},\
B^{a},\ 0\big),
\end{equation*}%
and so does the integrand in the generating functional $\mathcal{Z}%
_{k}(J,\Phi ^{\ast })$ of Green's functions (introduced in \cite{Wett-Reu-2}%
, \cite{LS} with the obvious change of the notation $\Phi _{A}^{\ast }\equiv
K_{A}$) for the vanishing sources, $J_{A}=$ $\bigl(J_{\mu }^{a},J_{C}^{a},J_{\bar{C%
}}^{a},J_{B}^{a}\bigr)(x)=$ $0$, and the regulator action, $S_{k}=0$,
\beq
\label{ZWk}
{\cal Z}_k(J,\Phi^*) =
\int d\Phi
\exp \Big\{\frac{i}{\hbar}\big[
S_{FP}(\Phi)+\Phi^*_A {s}\Phi^A + S_k(\Phi)+J\Phi\big]\Big\} \ = \exp\{\sfrac{\im}{\hbar}\mathcal{W}_k(J,\Phi^*) \}\,.
\eeq
Before taking the limit $k\rightarrow 0$, the integrand in the case $J=0$
is not BRST-invariant, due to the easily verified inequality
\begin{equation*}
\delta _{\mu }\,S_{k}(\Phi )\,\neq \,0\,,
\end{equation*}%
whereas in the limit $k\rightarrow 0$ the functionals $\mathcal{Z}_{k}$, $%
\mathcal{W}_{k}$ take correct values, identical with the usual
generating functionals $Z,W$. The  average  effective action $\Gamma
_{k}=\Gamma _{k}(\Phi ,\Phi ^{\ast })$, being the generating functional
of vertex functions in the presence of regulators, is introduced
according to the rule described by Eq. (\ref{EAq}) in
Section~\ref{gengd},
\begin{equation} \label{EAk}
\Gamma _{k}(\Phi ,\Phi ^{\ast })=\mathcal{W}_{k}(J,\Phi ^{\ast })-J\Phi
\,,\qquad \Phi ^{A}=\frac{\delta \mathcal{W}_{k}}{\delta J_{A}}\,,
\end{equation}%
with the obvious consequences of the Legendre transformation (\ref{EAk}), $J_A = (\delta \Gamma_k)/(\delta \Phi^A)$.
Note, first of all, that the average effective action, by analogy with Eq. (\ref%
{GMEq-loop}), satisfies an equation and possess tree-level and one-loop
approximations which are similar to those for $\Gamma _{M}$ in Eqs. (\ref{0loop}%
), (\ref{1loop}), however, with $S_{k}$, instead of the functional $M$.
Second, as to the regulator functions, we suppose that they model the
non-perturbative contributions to the self-energy part of the Feynman
diagrams, so that the dependence on the parameter $k$ enables one to extract
some additional information about the scale dependence of the theory beyond
the loop expansion \cite{Polch}. Third, the Ward (Slavnov--Tailor)
identities \cite{Slav}, \cite{Tay}
 for the functionals $\mathcal{Z}_{k},\mathcal{W}_{k}$ and $\Gamma
_{k}$ are easily obtained from the general results (\ref{wardZM}), (\ref%
{WIWBV}) and (\ref{WIGammaBV}) for ${Z}_{M},{W}_{M}$ (\ref{ZSfull}) and $%
\Gamma _{M}$ (\ref{EAq}), and, due to the property $M^{\ast
A}\equiv S_{k}^{\ast
A}\equiv 0$, take the form:

For $\mathcal{Z}_{k}$,%
\beq\label{wardzk}
J_A\frac{\delta {\cal Z}_k}{\delta \Phi^*_A} \,+\,
\frac{\hbar}{\im}\int d^Dx \Big[
(R_{k,A})^{ab}_{\mu\nu}
\,\frac{\delta^2{\cal Z}_k}{\delta J^b_{\nu}\delta A^{*a}_{\mu}}
\,-\,
(R_{k,gh})^{ab}\,\frac{\delta^2 {\cal Z}_k}{\delta J_C^b\delta {\bar C}^{*a}}
\,+\,
(R_{k,gh})^{ab}\,\frac{\delta^2 {\cal Z}_k}{\delta J_{\bar{C}}^a\delta C^{*b}}\Big] = 0\,,
\eeq
for $\mathcal{W}_{k}$,%
\begin{eqnarray}
&&J_{A}\frac{\delta \mathcal{W}_{k}}{\delta \Phi _{A}^{\ast }}\,+\,\frac{%
\hbar }{\text{i}}\int d^{D}x\Big[(R_{k,A})_{\mu \nu }^{ab}\,\frac{\delta ^{2}%
\mathcal{W}_{k}}{\delta J_{\nu }^{b}\delta A_{\mu }^{\ast a}}%
\,-\,(R_{k,gh})^{ab}\,\frac{\delta ^{2}\mathcal{W}_{k}}{\delta
J_{C}^{b}\delta {\bar{C}}^{\ast a}}\,+\,(R_{k,gh})^{ab}\,\frac{\delta ^{2}%
\mathcal{W}_{k}}{\delta J_{\bar{C}}^{a}\delta C^{\ast b}}  \notag \\
&&+\frac{\text{i}}{\hbar }\Big\{(R_{k,A})_{\mu \nu }^{ab}\,\frac{\delta
\mathcal{W}_{k}}{\delta J_{\nu }^{b}}\,\frac{\delta \mathcal{W}_{k}}{\delta
A_{\mu }^{\ast a}}\,-\,(R_{k,gh})^{ab}\,\frac{\delta \mathcal{W}_{k}}{\delta
J_{C}^{b}}\,\frac{\delta \mathcal{W}_{k}}{\delta {\bar{C}}^{\ast a}}%
\,+\,(R_{k,gh})^{ab}\,\frac{\delta \mathcal{W}_{k}}{\delta J_{\bar{C}}^{a}}%
\frac{\delta \mathcal{W}_{k}}{\delta C^{\ast b}}\Big\}\Big]\ =\ 0,
\label{WIWkJ}
\end{eqnarray}%
for $\Gamma _{k}$,%
\begin{eqnarray}
&&\frac{1}{2}\bigl(\Gamma _{k}\,,\Gamma _{k}\bigr)-\int d^{D}x\Big[%
(R_{k,A})_{\mu \nu }^{ab}\,A^{b\nu }\,\frac{\delta \Gamma _{k}}{\delta
A_{\mu }^{\ast a}}\,-\,(R_{k,gh})^{ab}\,C^{b}\,\frac{\delta \Gamma _{k}}{%
\delta {\bar{C}}^{\ast a}}\,+\,(R_{k,gh})^{ab}\,{\bar{C}}^{a}\,\frac{\delta
\Gamma _{k}}{\delta {C}^{\ast b}}  \notag \\
&&+i\hbar \Big\{(R_{k,A})_{\mu \nu }^{ab}\,\big(\Gamma _{k}^{^{\prime \prime
}-1}\big)^{b\nu \,A}\,\frac{\delta _{l}^{2}\Gamma _{k}}{\delta \Phi
^{A}\,\delta \Phi _{\mu }^{\ast a}}-(R_{k,gh})^{ab}\,\big(\Gamma
_{k}^{^{\prime \prime }-1}\big)^{{b}A}\,\frac{\delta _{l}^{2}\Gamma _{k}}{%
\delta \Phi ^{A}\,\delta {\bar{C}}^{\ast a}}  \notag \\
&&\qquad +\,(R_{k,gh})^{ab}\,\big(\Gamma _{k}^{^{\prime \prime }-1}\big)%
^{aA}\,\frac{\delta _{l}^{2}\Gamma _{k}}{\delta \Phi ^{A}\,\delta {C}^{\ast
b}}\Big\}\Big]=0\,.  \label{STGk}
\end{eqnarray}%
The supermatrix $(\Gamma _{k}^{^{\prime \prime }-1})$ is the inverse of
$\Gamma _{k}^{^{\prime \prime }}$, with the elements determined
by analogy with Eqs. (\ref{varGBV}), (\ref{invGBV}), with the obvious
replacement $\Gamma \rightarrow \Gamma _{k}$. In the limit \ $k\rightarrow 0$%
, the identities (\ref{wardzk}), (\ref{WIWkJ}), (\ref{STGk}) are reduced to
the standard Ward identities (\ref{WIBV}).

The consistency of the FRG method, based on the introduction of Eq. (\ref{regSk}%
), means that the values of the average effective  actions $\Gamma _{k}$
calculated for two different gauges determined by $\chi ^{a}$ and $\chi ^{a}+%
\overline{\Delta }\chi ^{a}$ corresponding, in view of Eq. (\ref{psiGk}), to
the gauge functionals $\Psi $ and $\Psi +\overline{\Delta }\Psi $, should
coincide on the mass-shell for any value of the parameter $k$ (i.e., along
the FRG trajectory, but not only in its boundary points).
For completeness, let us recall that the FRG flow equation
for $\Gamma_{k}$, which describes the FRG trajectory, reads \cite{LS}
as follows, with account taken of the notation $\partial_t = k\frac{d}{dk}$:
\beq
\label{FRGeq}
\partial_t\Gamma_k \,=\, \partial_t S_k
\,-\, \frac{\hbar}{\im}\,\,\Big\{
\frac{1}{2}\,\partial_t(R_{k,A})^{ab}_{\mu\nu}\,
\big(\Gamma^{''-1}_k\big)^{(a\mu)(b\nu)}
\,+\,\partial_t(R_{k,gh})^{ab}\,\big(\Ga^{''-1}_k\big)^{ab}\Big\}\,,
\eeq
which has the same form for the $\Phi^*_A$-independent part of $\Gamma_k$,
due to the parametric dependence on $\Phi^*_A$ of all the terms in (\ref{FRGeq}).

Due to the result (\ref{Mdelta1r}) for a finite variation of the BRST symmetry breaking
term, the variation of the regulator action $S_{k}$ under the variation of
the gauge condition has the form
\beq\label{Skdelta1r}
 \overline{\Delta} S_k \ =\ S_{k,A}S^A_\Psi \overline{\Delta}\Psi \big(s\overline{\Delta}\Psi\big)^{-1}\Big[\exp \left\{-\frac{i}{\hbar} s\big(\overline{\Delta}\Psi\big)\right\} -1\Big],
\eeq
with the use of the Slavnov operator $s$. The corresponding gauged BRST
transformation, leading to the variation of $S_{k}$, as applied to the
generating functional $\mathcal{Z}_{k}$ in the gauge $\Psi (\Phi )$ (\ref%
{psiGk}), must be characterized by the parameter $\Lambda (\Phi )$ given by
\begin{equation}\label{LamPhisk}
{\Lambda }(\Phi )\ =\ \overline{\Delta }\Psi (\Phi )\big(s\overline{\Delta }%
\Psi (\Phi )\big)^{-1}\Big[\exp \left\{ -\frac{i}{\hbar }s\big(\overline{%
\Delta }\Psi (\Phi )\big)\right\} -1\Big].
\end{equation}%
According to Eq. (\ref{varGammaM}), in the case under consideration a finite
variation of the average effective action $\Gamma _{k}$  with allowance
for the explicit form of $\overline{\Delta }S_k$ (\ref{Skdelta1r}) takes the form
\begin{eqnarray}
\overline{\Delta }\Gamma _{k} &=& \exp \Big\{\frac{\text{i}}{\hbar }\langle\overline{\Delta }S_k\rangle%
\Big\} \bigg(\langle\overline{\Delta }%
S_{k{}A}\rangle
\sum_{n\geq 1}\frac{1}{n!}\Big\{ \langle\overline{\Delta }\Psi_B\rangle
\textstyle\Big(- {\widehat F}^B +\frac{\text{i} }{\hbar}\frac{\delta \Gamma_k}{\delta \Phi _{B}^{\ast }}  %
\Big)\Big\}^{n-1} \textstyle \Big(- {\widehat F}^A +\frac{\text{i} }{\hbar}\frac{\delta \Gamma_k}{\delta \Phi _{A}^{\ast }}  %
\Big)\nonumber \\
&&   + \sum_{n\geq 1}\frac{1}{n!}\Big\{ \langle\overline{\Delta }\Psi_B\rangle
\textstyle\Big(- {\widehat F}^B +\frac{\text{i} }{\hbar}\frac{\delta \Gamma_k}{\delta \Phi _{B}^{\ast }} %
\Big)\Big\}^{n-1}  \Big\{-\big(\Gamma _{k},\ \ \big)\ +\ \displaystyle{\widehat{S}}_{k{}A}\frac{\delta }{\delta \Phi _{A}^{\ast }}  \notag \\
&&  -\ \frac{\text{i}}{\hbar }\Big[{\widehat{S}}_{k{}A}\frac{\delta \Gamma _{k}}{%
\delta \Phi _{A}^{\ast }},\ \Phi ^{B}%
\Big]\frac{\delta _{\mathit{l}}}{\delta \Phi ^{B}}\, \Big\}
\bigg)\langle \overline{%
\Delta }\Psi \rangle
  +  \sum_{n\geq 1}\frac{1}{n!}\left(\frac{i}{\hbar}\right)^{n-1}\langle\overline{\Delta }S_k\rangle
,  \label{varGammak}
\end{eqnarray}%
where (\ref{laradepsi}) has been taken into account, and the operator ${\widehat{F}}^{A}$ is
now determined as%
\begin{equation}\label{FAsk}
{\widehat{F}}^{A}=-\frac{\delta }{\delta \Phi _{A}^{\ast }}\ +\
(-1)^{\varepsilon _{B}(\varepsilon _{A}+1)}(\Gamma _{k}^{^{\prime \prime
}-1})^{BC}\Big(\frac{\delta _{\mathit{l}}}{\delta \Phi ^{C}}\frac{\delta
\Gamma _{k}}{\delta \Phi _{A}^{\ast }}\Big)\frac{\delta _{\mathit{l}}}{%
\delta \Phi ^{B}}\ .
\end{equation}%
Being linear in $\langle \overline{\Delta }\Psi \rangle $ and,
due to (\ref{Skdelta1r}),
also in $\langle \overline{\Delta }S_{k}\rangle $, the variation $\overline{%
\Delta }\Gamma _{k}(J,\Phi ^{\ast })$ takes another
\textquotedblleft local-like\textquotedblright\ form; see Eq.
(5.6) in \cite{LS}:
\beq\overline{\Delta}\Gamma_k \ =\
-\big(\Gamma_k,\langle\overline{\Delta}\Psi\rangle\big)\ +\
{\widehat S}_{k{}A}\frac{\delta}{\delta \Phi^{*}_{A}} \langle\overline{\Delta}\Psi\rangle  -\ \frac{\im}{\hbar}\Big[{\widehat S}_{k{}A} \frac
{\delta\Gamma_k}{\delta\Phi^*_A},\
\Phi^B\Big] \frac{\delta_{\it
l}}{\delta\Phi^B}\,\langle\overline{\Delta}\Psi\rangle+\ \langle
\overline{\Delta }S_{k}\rangle \label{varGklin}  . \eeq
There is a representation equivalent to Eq. (\ref{varGklin}) and similar to Eqs. (\ref%
{varGammaFlin}), (\ref{FAdef}):%
\begin{equation}\label{varGkfin}
\overline{\Delta }\Gamma _{k}={\frac{\delta \Gamma _{k}}{\delta \Phi ^{A}}%
}{\widehat{F}}^{A}\,\langle \overline{\Delta }\Psi \rangle \ -\ {\widehat{S}}%
_{k{}A}{\widehat{F}}^{A}\langle \overline{\Delta }\Psi \rangle \ +\ \langle
\overline{\Delta }S_{k}\rangle \ .
\end{equation}%
Due to the statement proved at the end of Section~\ref{gengd} [see Eq. (\ref%
{finvarGM})] regarding the presence in the gauge theory\ of a
soft BRST breaking term, we can state that the average effective action $%
\Gamma _{k}$, at least in the non-renormalized case, being evaluated
at its extremals, does not depend on the choice of the gauge
condition:
\beq\label{finvarGk} \frac{\delta\Gamma_k}{\delta\Phi^A}=0\texttt{ and } \langle\overline{\Delta}S_k\rangle  ={\widehat S}_{k{}A}{\widehat F}^A \langle\overline{\Delta}\Psi\rangle
\longrightarrow\qquad \overline{\Delta}\Gamma_k = 0\ ,
\eeq
provided that, in the approximation being linear with respect to
$\overline{\Delta }\Psi $, the variation of the regulators $S_{k}$
(\ref{Skdelta1r}) takes the form
\beq
\label{Skdelta1rlin}
\overline{\Delta} S_k \ =\ - s (S_{k}) \frac{\im}{\hbar}\overline{\Delta}\Psi\texttt{ and }{\Lambda}(\Phi)\ =\ -\frac{\im}{\hbar}\overline{\Delta}\Psi(\Phi) + o(\overline{\Delta}\Psi(\Phi)) ,
\eeq
which, after averaging with respect to the mean fields $\Phi ^{A}$
by using $\Gamma _{k}$, leads to%
\beq\label{repdSK}
\langle\overline{\Delta} S_k\rangle + \langle s (S_{k}) \frac{\im}{\hbar}\overline{\Delta}\Psi\rangle = 0  \Longleftrightarrow  \langle\overline{\Delta}S_k\rangle - {\widehat S}_{k{}A}{\widehat F}^A \langle\overline{\Delta}\Psi\rangle =0 .
\eeq
The result given by Eq. (\ref{finvarGk}) allows one to revise (in comparison with \cite{LS})
the statement
on the gauge-dependence of the average effective action, and therefore also on
the consistency of its introduction within the Lagrangian quantization
scheme for any value of the parameter $k$.
Indeed, the gauge dependence of the vacuum functional $Z_{k,\chi}$
and of the average effective action  $\Gamma_k$ on its extremals \cite{LS}
was explicitly shown respectively in (4.12) and (5.9) therein.
At the same time, the gauge independence of the average effective
action in  \cite{LS} was achieved on the mass-shell determined
in a larger space of fields  (see (6.31), (6.32) therein)
with additional degrees of freedom, by means of considering
the regulators $S_{k}$ as composite fields, however, without
taking into account the change of the regulators $S_{k}$
under a change of the gauge condition in (6.22), (6.27), (6.30), (6.31).
In this connection, note that the consideration
of the regulators as the composite fields following to approach
\cite{Reshetnyak2} -- where their change under a variation
of the gauge condition should be taken into account --
allows one to provide the gauge independence of  $\Gamma_k$
on the mass shell determined by the usual average fields $\Phi^A$ only.

Let us calculate the form of the regulator terms $S_{k}$ in different, but
mutually related gauges, setting as $S_{k}^{0}$ the values in a fixed gauge;
for the sake of definiteness, in the  Landau gauge. To this end,
let us consider a family of linear gauges given by the equation
\begin{eqnarray}
\chi ^{a}(A,B) &=&\Lambda _{\mu }(\partial ,\alpha ,\beta ,n)A^{\mu a}+\frac{%
\xi }{2}B^{a}=0 \ \ \mathrm{ with } \ \
\Lambda _{\mu }(\partial ,\alpha ,\beta ,n) = \alpha \partial _{\mu }+\beta
\frac{\kappa_{\mu\nu}}{n^2}n^{\nu } . \label{gengauge}
\end{eqnarray}%
Here, we have three numeric, $\alpha ,\beta ,\xi $, and one vector, $n^{\mu }$,
gauge parameters. From  $\alpha ,\beta ,\xi $, we can keep only two
numbers, $\beta ,\xi $, for instance, dividing
$\chi ^{a}(A,B)$ by $\alpha $.

Particular cases of these gauges can be obtained from the general
many-parameter family under the choices
\begin{eqnarray}
\alpha =1,\beta =0 &\rightarrow &\ \texttt{ family of  }R_{\xi -}\texttt{gauges},
\label{lingage1} \\
\beta = - \alpha , \kappa_{\mu\nu} = n^{\rho }\partial _{\rho}\eta_{\mu\nu},  n^{2}<0,\xi =0 &\rightarrow &\texttt{ generalized  Coulomb
gauges},  \label{lingage2} \\
\alpha =0,\  \kappa_{\mu\nu} = \eta_{\mu d-1}n_\nu, \xi =0 &\rightarrow &\texttt{ generalized axial gauges}.
\label{lingage3}
\end{eqnarray}%
The Landau and Feynman  gauges are obtained from the first family for the
respective choices $\xi =0$ and $\xi =1$. The usual Coulomb, $\chi
_{C}^{a}(A,B)$, and axial, $\chi _{A}^{a}(A,B)$, gauges are derived from the
second and third families by setting, $n^{\mu }=(1,0,...,0)$ and $%
n^{\mu }=(0,...,0,1)$ for the respective parameters. For
completeness, we have
\begin{eqnarray}
\chi _{C}^{a}(A,B) &=&\partial _{i}A^{ia}=0,\ \mathrm{\ \ \ for\ \ \ }\ \mu
=(0,i),  \label{Cgauge} \\
\chi _{A}^{a}(A,B) &=&A^{d-1{}{}a}=0.  \label{Agauge}
\end{eqnarray}

Denoting the Landau gauge as $\chi ^{a}(A,B)\big\vert_{\alpha
=1,\beta =\xi =0}\equiv \chi ^{a}(A)$, we can examine the form of
the regulators which arises for arbitrary values of the parameters
$\alpha ,\beta ,\xi $, $n^{\mu }$. Following Eq.
(\ref{Skdelta1r}), we immediately obtain the variation of
the gauge fermion and its Slavnov variation, respectively, %
\begin{eqnarray}
&&\overline{\Delta }\Psi  = \bar{C}{}^{a}\big(\chi ^{a}(A,B)-\chi ^{a}(A)%
\big)\ =\int d^{D}x\bar{C}{}^{a}\big(\{(\alpha -1)\partial _{\mu }+\beta
\frac{\kappa_{\mu\nu}}{n^2}n^{\nu }\}A^{\mu a}+\frac{\xi }{2}B^{a}%
\big),  \label{deltapsiYM} \\
&&s\overline{\Delta }\Psi  =  \int d^{D}x\Big\{B^{a}\big(\{(\alpha
-1)\partial _{\mu }+\beta \frac{\kappa_{\mu\nu}}{n^2}n^{\nu }\}A^{\mu a}+\frac{\xi }{2}B^{a}\big)  \notag \\
&&\phantom{s\overline{\Delta }\Psi  =} +\bar{C}{}^{a}\big((\alpha -1)\partial
_{\mu }+\beta \frac{\kappa_{\mu\nu}}{n^2}n^{\nu }\big)D^{\mu
ab}C^{b}\Big\}.  \label{sdeltapsiYM}
\end{eqnarray}%
so that the expression for $S_{k}=S_{k}^{0}+\overline{\Delta
}S_{k}$ reads
\begin{eqnarray}
S_{k} &=&S_{k}^{0}+\int d^{D}x\Big\{A^{a\mu }(x)(R_{k,A})_{\mu \nu
}^{ab}(x)D^{bc\nu }C^{c}+(R_{k,gh})^{ab}(x)\big(\textstyle\frac{1}{2}f^{bcd}{\bar{C}}%
^{a}C^{c}C^{d}-C^{b}B^{a}\big)\Big\}  \notag   \\
&&\times \overline{\Delta }\Psi \Big(s\overline{\Delta }\Psi \Big)^{-1}\Big[%
\exp \left\{ -\frac{i}{\hbar }s\big(\overline{\Delta }\Psi \big)\right\} -1%
\Big].\label{deltask}
\end{eqnarray}%
From Eqs. (\ref{deltapsiYM})--(\ref{deltask}), we find an
approximation linear in $\overline{\Delta }\Psi $,%
\begin{eqnarray}
S_{k}(\Phi ) &=&\int d^{D}x\Big\{\frac{1}{2}A^{a\mu }(x)(R_{k,A})_{\mu \nu
}^{ab}(x)A^{b\nu }(x)+{\bar{C}}^{a}(x)(R_{k,gh})^{ab}(x)C^{b}(x)\Big\}
\notag   \\
&&-\frac{\text{i}}{\hbar }\int d^{D}x\Big\{A^{a\mu }(x)(R_{k,A})_{\mu \nu
}^{ab}(x)D^{bc\nu }C^{c}+(R_{k,gh})^{ab}(x)\big(\textstyle\frac{1}{2}f^{bcd}{\bar{C}}%
^{a}C^{c}C^{d}-C^{b}B^{a}\big)\Big\}  \notag \\
&&\times \int d^{D}y\bar{C}{}^{e}(y)\Big\{\big((\alpha -1)\partial _{\rho
}+\beta \frac{\kappa_{\rho\nu}}{n^2}n^{\nu }\big)A^{\rho e}(y)+%
\frac{\xi }{2}B^{e}(y)\Big\},\label{deltasklin}
\end{eqnarray}%
depending now on all field variables and having the standard limit $%
S_{k}\rightarrow 0$ as the momentum-shell parameter tends to zero, $%
k\rightarrow 0$. For $\alpha {=}1,\beta {=}\xi {=}0$, the regulators $%
S_{k}(\Phi )$ are smoothly reduced to the initial ones
$S_{k}^{0}(\Phi )$, given in the Landau gauge, whereas the
expressions for $S_{k}(\Phi )$ in any gauges described by
Eqs. (\ref{lingage1})--(\ref{Agauge}) can now be explicitly obtained
from Eq. (\ref{deltasklin}).

In order to obtain the form of $S_{k}$ (\ref{deltasklin}) without the terms $%
\frac{\text{i}}{\hbar }$, so that this functional should start from the
tree-level term, we have to perform integration with respect to the
Faddeev--Popov ghost fields in the functional integral $\mathcal{Z}_{k}$ (%
\ref{ZWk}), and then extract the Faddeev-Popov operator (\ref{FPoper}), $
K^{ab}$, in the resulting gauge and exponentiate it with help of
the same Faddeev--Popov ghost fields.

It is interesting to investigate the consequences of the study of
gauge-dependence in the case of the Pauli--Villars regularization \cite{PV},
which does not preserve the gauge and therefore also the BRST invariance of the
regularized quantum action in the regularization
scheme without higher derivatives introduced in \cite{FaddeevSlavnov},
but we leave this study outside this paper's scope.

\section{Standard and Refined Gribov--Zwanziger Actions in Many-
\newline parameter Family of Gauges}\label{rGZ}

\noindent In this section, we apply the above general
consideration developed in Section~\ref{gengd} and adopted to the
case of the average effective action for Yang--Mills theories in Section~%
\ref{FRGandPV} in the case of the so-called Gribov--Zwanziger \cite{Zwanziger1}%
, \cite{Zwanziger2} and refined Gribov--Zwanziger theories, introduced in \cite%
{0806.0348} and examined in \cite{0808.0893}, \cite{0906.4257}, \cite%
{1102.0574}, \cite{1105.3371}. Let us remind that the
Gribov--Zwanziger theory is determined by the Gribov--Zwanziger
action $S_{GZ}(\Phi)$, given in the Landau gauge $\chi^a(A)=0$,
\begin{eqnarray}  \label{GZact}
S_{GZ}(\Phi)\=S_{FP}(\Phi)\ +\ M(A)\ ,
\end{eqnarray}
which contains an additive non-local BRST-non-invariant summand,
implying an inclusion of the Gribov horizon  \cite{Gribov} and
known as the Gribov horizon functional $M(A)$, with suppressed
continuous space-time coordinates $x,y$,
\begin{eqnarray}  \label{FuncM}
M(A)\=\gamma^2\,\big(f^{abc}A^b_{\mu}(K^{-1})^{ad}f^{dec} A^{e\mu}\ +\ D(N^2{%
-}1)\big)\ ,\texttt{ for } (K^{-1})^{ad}(K)^{db}=\delta^{ab},
\end{eqnarray}
which is determined by means of the Faddeev-Popov operator $(K)^{ab} = \partial_\mu D^{%
\mu{}ab}$ and the so-called thermodynamic (Gribov)
parameter~$\gamma$, introduced
in a self-consistent way by the gap equation \cite{Zwanziger1}, \cite%
{Zwanziger2}, \cite{Sorellas}
\begin{equation}
\frac{\partial }{\partial \gamma}\left( \frac{\hbar}{\text{i}} \,\mbox{ln}\, %
\Big[ \int\!D\Phi\ \exp\Big\{\frac{\text{i}}{\hbar}S_{GZ}(\Phi)\Big\}\Big]%
\right)= \frac{\partial \mathcal{E}_{vac}}{\partial \gamma} \= 0\
. \label{gapeq}
\end{equation}
In Eq. (\ref{gapeq}), we have used the definition of the vacuum energy $\mathcal{%
E}_{vac}$.  The idea to improve the Gribov--Zwanziger theory is
due to the facts that, in the first place, it fails to
eliminate all Gribov's copies, and, second, a non-zero value for
the Gribov parameter $\gamma$ is a manifestation of nontrivial
properties of the vacuum \cite{1102.0574} of the theory as a
consequence of restrictions on the Gribov horizon. The latter
means that there exist additional reasons for
non-perturbative effects, which can be encoded in a set of
dimension-2 condensate, $\langle A^{\mu a}A_\mu^a \rangle $, in
the case of a non-local Gribov--Zwanziger action with the
Yang--Mills gauge fields $A^{\mu a}$ only, as well as
in a similar set of dimension-2 condensates, $%
\langle A^{\mu a}A_\mu^a \rangle$, $\langle \bar{\varphi}^{\mu{}
ab}\varphi_\mu^{ab} \rangle{-}\langle \bar{\omega}^{\mu{}
ab}\omega_\mu^{ab} \rangle$, for a local Gribov--Zwanziger action,
$S_{GZ}(\Phi, \phi)$, with an equivalent local representation for
the horizon functional in terms of the functional $S_\gamma$,
given in an extended configuration space with auxiliary variables,
$\phi^{\bar{A}}$,
\begin{eqnarray}  \label{Sgamma}
 S_{GZ}(\Phi,\phi) &= & S_{FP}(\Phi)\ +\ S_{\gamma}(A,\phi)\ \  \mathrm{ with } \\
 S_{\gamma}&= & {\bar \varphi}^{ac}_\mu K^{ab} \varphi^{\mu{} bc}
- {\bar \omega}^{ac}_\mu K^{ab} \omega^{bc}_{\mu} + f^{amb}
(\partial_{\nu}{\bar
\omega}^{ac}_{\mu}) (D^{mp}_{\nu} c^p)\varphi^{bc}_{\mu}  \notag \\
& &\  + \gamma\,f^{abc}A_\mu^{a}(\varphi_\mu^{bc}-\bar{\varphi}%
_\mu^{bc})-D(N^2-1)\gamma^2.
\end{eqnarray}
Here, the fields $\phi^{\bar{A}}$ contain tensors being antisymmetric with
respect to the $su(N)$ indices,
\begin{eqnarray}
\bigl\{\phi^{\bar{A}}\bigr\}\= \bigl\{\varphi^{ac}_\mu\,,\,{\bar\varphi}%
^{ac}_\mu\,,\, \omega^{ac}_\mu\,,\,{\bar\omega}^{ac}_\mu\bigr\},
\end{eqnarray}
even for $\varphi^{ac}_\mu$, ${\bar\varphi}^{ac}_\mu$ (i.e., $%
\varepsilon(\varphi){=}\varepsilon(\bar{\varphi}){=}0$) and odd for $%
\omega^{ac}_\mu$, ${\bar\omega}^{ac}_\mu$ ($\varepsilon(\omega){=}%
\varepsilon(\bar{\omega}){=}1$), which form BRST
doublets~\cite{DudalSV},
\begin{eqnarray}
\delta_{\mu} \Big(\varphi^{ac}_\nu, {\bar\varphi}^{ac}_\nu\Big ) &= \Big(%
\omega^{ac}_\nu , 0\Big)\mu\qquad\ \, \delta_{\mu}\Big(\omega^{ac}_\nu, {%
\bar\omega}^{ac}_\nu\Big) = \Big(0,{\bar\varphi}^{ac}_\nu\Big)\mu
.
\end{eqnarray}
Both the non-local $M(A)$ and local $S_{\gamma}$ horizon functionals
are not BRST-invariant:
\beq
\label{sM}
&& \hspace{-0.5em}sM \=\gamma^2f^{abc}f^{cde}\bigl[2D^{bq}_{\mu}C^q(K^{-1})^{ad}-
f^{mpn}A^b_{\mu}(K^{-1})^{am}K^{pq}C^q(K^{-1})^{nd}\bigr]A^{e\mu}\ \neq\ 0,\\
\label{varSg}
&& \hspace{-0.5em} sS_{\gamma}\= \gamma f^{adb}\big[
\big( D^{de}_{\mu}C^e(\varphi^{\mu ab}{-}{\bar\varphi}^{\mu ab})+
A^d_{\mu}\omega^{\mu ab}\big)\big]\ \neq\ 0\ ,
\eeq
where account has been taken of the relation
$s K^{ab}  \ =\ f^{acb}K^{cd}  C^{d}$,
with \textbf{the latter Slavnov variation}, together with the
representation for $S_\gamma$, being different from those of
\cite{ll2}. The problem of finding the Gribov horizon functional
in reference frames other than the Landau gauge has been
considered in various papers. In \cite{SS}, this problem was first
solved in the approximation being quadratic in the fields for the
linear covariant $R_\xi$-gauges given by Eqs. (\ref{gengauge}),
(\ref{lingage1}) for a small value of the parameter $\xi$; another
form of the functional $M(A,\xi)$ was suggested in \cite{rl}, and
also with the help of the gauged (field-dependent) BRST
transformations in the recent paper \cite{ll2}. Of course, the
suggested result requires a verification of the fact that the
functional derived actually satisfies the requirement that it
should single out the first Gribov horizon region for the gauge fields
$A^{\mu{}a}$ in the $R_\xi$-gauge, because an extraction of this
region via the functional $M(A)$ was determined non-perturbatively
\cite{Zwanziger1} in the Landau gauge only, whereas a
corresponding rigorous proof for $M(A,\xi)$, i.e., that it
actually provides the restriction for the gauge fields
$A^{\mu{}a}$
within the Gribov region $%
\Omega(\xi)$,
\begin{eqnarray}  \label{grrxi}
\Omega(\xi) = \Big\{A^{\mu{}a}\big\vert \chi ^{a}(A,B)\big\vert_{\alpha
=1,\beta =0}=0, K^{ab}(\xi)\geq 0 \Big\},
\end{eqnarray}
is absent  in the literature in an explicit way.

As we turn to the refined Gribov--Zwanziger theory, let us propose
the refined Gribov--Zwanziger action in a non-local form, and,
along the lines of \cite{0806.0348}, \cite{0808.0893}, \cite{0906.4257}, \cite{1102.0574},
\cite{1105.3371}, also in a local form, as follows:
\begin{eqnarray}
&& S_{GZ}(\Phi) \rightarrow S_{RGZ1}(\Phi)=S_{GZ}+\frac{m^2}{2} {A^a_{\mu}}{%
A^{\mu{}a}} \;,  \label{RGZn} \\
&& S_{GZ}(\Phi,\phi) \rightarrow S_{RGZ2}(\Phi,\phi)=S_{GZ}(\Phi,\phi)+\frac{m^2}{2} {%
A^a_{\mu}}{A^{\mu{}a}} - {M^2} \left(\overline\varphi_\mu^{ab} \varphi^{\mu{}%
ab}-\overline\omega_\mu^{ab}\omega^{\mu{}ab}\right),
\label{RGZ}
\end{eqnarray}
which can, of course, be considered as theories with composite
operators.

The only non-vanishing Slavnov variations are those of the first composite fields:
\begin{eqnarray}
s\left(\frac{m^2}{2} {A^a_{\mu}}{A^a_{\mu}}\right) \ = m^2{A^a_{\mu}}%
\partial^{\mu}C^a \ne 0,   \ \ \mathrm{whereas} \ \     s\big({M^2} \left(\overline\varphi_\mu^{ab} \varphi^{\mu{}%
ab}-\overline\omega_\mu^{ab}\omega^{\mu{}ab}\right)\big) \ = 0 \;,
\label{sRGZn}
\end{eqnarray}
so that the only new BRST-non-invariant term is $\frac{1}{2}m^2 {A^a_{\mu}}{A^{%
\mu{}a}}$.

%%%%%%%%%%%%%%%%%%%%%%%%%%%%%%%%$$$%%%%%%%%%%%%%%%%%%%%%%%%%%%%%%%%%%%%%%%%%%%%%%%%%

By virtue of the properties (\ref{sM}), (\ref{varSg}) of the functionals $M(A)$ and $%
S_\gamma$, as well as due to Eq. (\ref{sRGZn}) with the
composite fields $M+ \frac{1}{2} m^2{A^a_{\mu}}{A^{\mu{}a}}$,
and $S_\gamma+ \frac{1}{2} m^2{A^a_{\mu}}{A^{\mu{}a}}$ + ${M^2} (\overline\varphi \varphi%
-\overline\omega\omega)$, in Eq.~(\ref{FuncM}), these functionals
trivially satisfy both the quantum~~(\ref{brbrst}) and
classical~~(\ref {brbrstclas}) conditions of soft BRST symmetry breaking,
because of the independence on antifields.

To establish the gauge-independence of physical quantities in
these theories, we have to examine the models in various
gauges from the many-parameter family (\ref{gengauge}),
thus explicitly extending the result of \cite%
{ll2}. In this case, the Faddeev--Popov action is written as
follows:
\begin{eqnarray}\label{FPagen}
S_{FP}({\Phi,\alpha,\beta, n^\mu, \xi})= S_0(A) + {\bar C}^a
\Lambda^\mu(\partial,\alpha,\beta,n) D_\mu^{ab}C^b  +
\Lambda_\mu(\partial,\alpha,\beta,n) A^{\mu a} B^a  +  {\textstyle\frac{\xi%
}{2}} B^a B^a\ .
\end{eqnarray}
The Faddeev--Popov operator $K^{ab}=\Lambda^\mu D_\mu^{ab}$ depends on $(\alpha,\beta,n)$, but not on~$%
\xi$, and the functional $M$ should be removed from $%
(\alpha,\beta,n,\xi){=}(1,0,n,0)$. However, since $K^{ab}$ cannot be
Hermitian~\cite{SS}, \cite{rl} the application of the Zwanziger trick developed in the Landau  gauge seems to be impossible.
Now, we apply the result of the preceding Sections~\ref{gengd},~\ref%
{FRGandPV} to gauged BRST transformations, and then, following Eqs.  (\ref{Mdelta1r}), (%
\ref{Skdelta1r}), the variation of the gauge fermion $\overline{\Delta}%
\Psi$ and its Slavnov variation $s\overline{\Delta}\Psi$ are given
by Eqs. (\ref{deltapsiYM}), (\ref{sdeltapsiYM}), so that the form
of the Gribov horizon functional $M(\Phi,\alpha,\beta,n,\xi)\equiv
\tilde{M}$ in the gauge under consideration reads, $\tilde{M} = M
+ \overline{\Delta} M$,
\begin{eqnarray}  \label{deltaM}
\tilde{M} & = & M(A) + \gamma^2f^{abc}f^{cde}\bigl[%
2D^{bq}_{\mu}C^q(K^{-1})^{ad}-
f^{mpn}A^b_{\mu}(K^{-1})^{am}K^{pq}C^q(K^{-1})^{nd}\bigr]A^{e\mu}  \notag \\
&& \times \overline{\Delta}\Psi \Big(s\overline{\Delta}\Psi\Big)^{-1}\Big[%
\exp \left\{-\frac{i}{\hbar} s\big(\overline{\Delta}\Psi\big)\right\} -1%
\Big] .
\end{eqnarray}
In the liner approximation with respect to $\overline{\Delta}\Psi$, we have
\begin{eqnarray}  \label{deltaMlin}
\tilde{M} & = & M(A) -\frac{\imath}{\hbar} \gamma^2f^{abc}f^{cde}\bigl[%
2D^{bq}_{\mu}C^q(K^{-1})^{ad} -
f^{mpn}A^b_{\mu}(K^{-1})^{am}K^{pq}C^q(K^{-1})^{nd}\bigr]A^{e\mu}  \notag \\
&& \times \bar{C}{}^h\Big\{\big((\alpha-1)\partial_\rho + \beta
 \frac{\kappa_{\rho\nu}}{n^2}n^{\nu }\big)A^{\rho h} + \frac{\xi}{2}B^h \Big\}.
\end{eqnarray}
For $\alpha{=}1, \beta{=}\xi{=}0$, the Gribov horizon functional ${M}%
(\Phi,\alpha,\beta,n,\xi) $ reduces smoothly to $M(A)$ given in
the Landau gauge, whereas the expressions for
${M}(\Phi,\alpha,\beta,n,\xi)$ in any linear gauges are now described
by Eqs. (\ref{lingage1})--(\ref{Agauge}). Thus, for $\alpha{=}1,
\beta{=}0,\xi{=}1$ we deduce from Eq. (\ref{deltaM}) the Gribov
horizon functional in the Feynman gauge as in \cite{ll2}, whereas
in the Coulomb gauge $\chi^a_C(A,B) = \partial_iA^{i{}a} = 0$,
obtained by setting $n^\mu = (1,0,0,0)\equiv n_0^\mu$, $\alpha =
\beta= 1,\xi = 0 $ in Eq. (\ref{gengauge}), in which the Gribov
copies were first discovered \cite{Gribov}, the functional
${M}(\Phi,1,1,n_0,0)\equiv M_C$ has the form
\begin{eqnarray}  \label{MCoulomb}
{M}_C & =& M(A) +
\gamma^2f^{abc}f^{cde}\bigl[2D^{bq}_{\mu}C^q(K^{-1})^{ad}-
f^{mpn}A^b_{\mu}(K^{-1})^{am}K^{pq}C^q(K^{-1})^{nd}\bigr] A^{e\mu}  \notag \\
&&\times \overline{\Delta}\Psi_C \Big(s\overline{\Delta}\Psi_C\Big)^{-1}\Big[%
\exp \left\{-\frac{i}{\hbar} s\big(\overline{\Delta}\Psi_C\big)\right\} -1%
\Big], \\
&&\texttt{\ for }\overline{\Delta}\Psi_C \=
\bar{C}{}^a\partial_0A^{0
a},\quad s \overline{\Delta}\Psi_C\= B^a\partial_0A^{0 a} + \bar{C}{}%
^a\partial_0D^{0 ab}C^b .\label{sdCoulomb}
\end{eqnarray}
For the linear $\gamma$-dependent part of the functional
$S_\gamma$, which is now BRST-non-invariant, examined in the
general gauge $\chi^a(A,B)$ from the family (\ref{gengauge}), we
have an expression similar to Eq. (\ref{deltaM}),
\begin{eqnarray}
\gamma\frac{\partial}{\partial\gamma}S_\gamma(\Phi,\phi,\alpha,\beta,n,\xi)
& = & \gamma\frac{\partial}{\partial\gamma}S_\gamma(1,0,n,0) + \gamma f^{adb}%
\big[ \big( D^{de}_{\mu}C^e(\varphi^{\mu ab}{-}{\bar\varphi}^{\mu
ab})+
A^d_{\mu}\omega^{\mu ab}\big)\big]  \notag \\
&& \times \overline{\Delta}\Psi \Big(s\overline{\Delta}\Psi\Big)^{-1}\Big[%
\exp \left\{-\frac{i}{\hbar} s\big(\overline{\Delta}\Psi\big)\right\} -1\Big]%
. \label{deltaSg}
\end{eqnarray}
On the other hand, in the Coulomb gauge we have the same expression for $\gamma\frac{%
\partial}{\partial\gamma}S_\gamma$, given by Eq. (\ref{deltaSg}), however, with $%
\overline{\Delta}\Psi_C, s\overline{\Delta}\Psi_C$ given by Eq. (\ref%
{sdCoulomb}). Finally, for the BRST-non-invariant term $\frac{m^2}{2} {A^a_{\mu}}%
{A^a_{\mu}}$, we have a presentation in the gauge
(\ref{gengauge}) with account taken of Eqs. (\ref{deltapsiYM}),
(\ref{sdeltapsiYM}),
\begin{eqnarray}  \label{mAAgg}
\frac{m^2}{2} {A^a_{\mu}}{A^a_{\mu}} \to \frac{m^2}{2} {A^a_{\mu}}{A^a_{\mu}}%
+ m^2{A^a_{\mu}}\partial^{\mu}C^a   \overline{\Delta}\Psi \Big(s\overline{\Delta}\Psi\Big)^{-1}\Big[%
\exp \left\{-\frac{i}{\hbar} s\big(\overline{\Delta}\Psi\big)\right\} -1\Big]%
,
\end{eqnarray}
and also in the Coulomb gauge,
\begin{eqnarray}  \label{mAACg}
\frac{m^2}{2} {A^a_{\mu}}{A^a_{\mu}} \to \frac{m^2}{2} {A^a_{\mu}}{A^a_{\mu}}%
+ m^2{A^a_{\mu}}\partial^{\mu}C^a  \overline{\Delta}\Psi_C \Big(s\overline{\Delta}\Psi_C\Big)^{-1}%
\Big[\exp \left\{-\frac{i}{\hbar}
s\big(\overline{\Delta}\Psi_C\big)\right\} -1\Big].
\end{eqnarray}
Summarizing, we state that the Gribov horizon
functional and the local functional $S_\gamma$ are now obtained
explicitly in an arbitrary gauge from the many-parameter family,\footnote{It is formally
possible to consider the Gribov horizon
functional in the axial gauge $\chi _{A}^{a}$ (\ref{Agauge})
following Eq.~(\ref{deltaM}); however, it is an algebraic gauge
without a space-time derivative, which ensures that there is
no problem of Gribov copies due to Singer's result \cite{Singer}.}
described by
Eq. (\ref{gengauge}), as well as the total Gribov--Zwanziger action
in its local and non-local forms. The same takes place for the
refined Gribov--Zwanziger action, which is the principal result of
this section. Note that the solution of this problem is based
entirely on the concept of gauged (field-dependent) BRST
transformations.

We can now revise our final statement of \cite{llr} and maintain
that the soft breaking of BRST symmetry is not in conflict with
the gauge-independence of physical quantities in Yang--Mills theories
with the Gribov horizon both
in the Gribov-Zwanziger and in refined Gribov-Zwanziger theories.

\section{Conclusion}\label{concl}

\noindent We have elaborated a treatment of general gauge theories
with arbitrary gauge-fixing in the presence of soft breaking
of the BRST symmetry in the field-antifield formalism. To this end, we
have studied the concept of gauged (equivalently,
field-dependent) BRST transformations for theories more general
than the Yang--Mills theory, and calculated the exact Jacobian (\ref{jacobianres})
of the corresponding change of variables in the path
integral determining the generating functionals of Green's
functions, including the effective action. We have argued,
on a basis of analyzing the non-linear functional equation (\ref{eqlambdapsi}) %
for an unknown field-dependent odd-valued parameter,
which we call the ``compensation equation'',\footnote{Note
that the term ``compensation equation'' has been recently suggested
\cite{BLThf}, \cite{BLTfin} for BRST symmetry in the study
of finite BRST--BFV and BRST--BV transformations, respectively,
as well as for BRST-antiBRST symmetry in Yang--Mills \cite{MRnew}
and general gauge theories in Lagrangian \cite{MRnew2}, \cite{MRnew3}
and  generalized Hamiltonian \cite{MRnew1}, \cite{BLThfext}
formulations.} that for
any finite change of the gauge condition $\Psi \to \Psi+
\overline{\Delta}\Psi$ there exists a gauged
BRST transformation with a field-dependent parameter $\Lambda(\Phi,\Phi^*|\overline{\Delta}\Psi)$
in (\ref{solcompeq}), depending on $\overline{\Delta}\Psi$, which
permits an entire compensation of the finite change of the vacuum
functional, i.e., $Z_{\Psi} = Z_{\Psi+\overline{\Delta}\Psi}$.

We have investigated the influence of BRST-non-invariant
terms, $M$, added to the quantum action constructed within the BV formalism
and satisfying the so-called soft BRST symmetry breaking condition, on
the properties of gauge-dependence of the corresponding effective
action $\Gamma_M$. To study this problem, we have, for the first
time, calculated  finite changes of the generating functionals $Z_M$, $W_M$
and the effective action
$\Gamma_M$ under a finite change of the gauge condition
(\ref{varZMf}), (\ref{varWMfin}), (\ref{varGammaM}) and found
that, at least with accuracy up to the linear terms in the variation
of the gauge-fixing functional
$\overline{\Delta}\Psi$, the effective action  does not depend on
its extremals on the choice of gauge, provided that the
change of the BRST-broken term is subject to
a corresponding gauged BRST transformation with the parameter
$\Lambda(\Phi,\Phi^*|\overline{\Delta}\Psi) $ determined
by (\ref{Mdeltalin}) and used in (\ref{finvarGM}),
which is our  principal result. Thereby,
the concept of soft BRST symmetry breaking does not violate the
consistency of Lagrangian quantization within the perturbation
theory, so that the suggested prescription allows one, first of all,
to obtain perturbatively the form of the soft BRST symmetry broken term in a
different gauge by means of Eq. (\ref{Mdeltalin})
[for a gauge theory of rank 1 with help of (\ref{Mdelta1r})], at least for
gauges being sufficiently close to each other, and, second, to
restore the gauge-independence of the effective action at its
extremals, and therefore also the gauge-independence of the conventional physical
$S$-matrix. We believe that these results should also be valid for
a renormalized theory with soft BRST symmetry breaking;
however, this requires a detailed proof.

We have demonstrated the applicability of our statements in the
case of the functional renormalization group approach to the
Yang--Mills and gravity theories and found, within the
many-parameter family of linear gauges (\ref{gengauge}),
the form of the regulator functionals in arbitrary (\ref{deltask}) and linear gauges (\ref{deltasklin})
from the same family, starting from those given, e.g., in the
Landau gauge. This construction allows one to restore
the gauge-independence of the average effective action $%
\Gamma_k$ along the entire trajectory of a FRG flow (\ref{finvarGk})
without having recourse to the composite fields technique.
Finally, the general concept of the gauged BRST transformations related to
the same gauge theory, however, given in different gauges, appears
to be very useful in constructing the Gribov--Zwanziger and the
refined Gribov--Zwanziger actions for a many-parameter family of
gauges, including the Coulomb, axial and covariant gauges (\ref{deltaMlin}), (\ref{mAACg}).
This result extends the Gribov--Zwanziger theory with $R_\xi $-gauges
examined in \cite{ll2}. At the same time, there arises a problem
of comparing the form of the horizon functional in the Coulomb gauge
obtained perturbatively by means of gauged BRST transformations
(\ref{MCoulomb}), (\ref{sdCoulomb}) with the horizon functional
obtained following to the Zwanziger non-perturbative recipe
\cite{HFZwanziger}, which is planned to consider as a separates study.
Of course, our arguments are valid for gauge theories with
soft breaking of the BRST symmetry in case the transformed BRST breaking
terms satisfy the same conditions in the final gauge as the untransformed
ones in the initial gauge, however, with a possible violation of
the condition (\ref{brbrst}) of soft BRST symmetry breaking.
For instance, this means that for the Gribov horizon functional
in a different gauge amongst the examined family of gauges
one needs to verify the validity of extracting the Gribov horizon
precisely from the configuration space of Yang--Mills fields,
perhaps with the examined dimension-2 condensate.

Finally, it may be hoped that, due to the appearance of the
Higgs field in view of the spontaneous breaking of the initial
gauge invariance related to the group $SU(2)$ for the electroweak
Lagrangian, one can examine an addition (associated with the Higgs field)
to the gauge-invariant (with respect to the $SU(2)$ group) action
of a soft BRST-breaking term, so that the description of the resulting
model will be made consistent in the conventional Lagrangian
path integral approach developed in this paper.
We consider this problem as the next one to be examined.

Concluding, let us mention, first, the treatment of the Gribov horizon
functional as a composite field \cite{Reshetnyak2}, second,
the recently obtained BRST-antiBRST extension \cite{MRnew}
of the Gribov--Zwanziger theory in different gauges in a way consistent
with the gauge independence of the physical $S$-matrix, third,
the concept of soft BRST-antiBRST symmetry breaking developed
on a basis of finite field-dependent BRST-antiBRST transformations
in \cite{MRnew3}.

\section*{Acknowledgments}

\noindent I thank V.A. Rubakov, S.V. Demidov and the participants
of the Seminar on Theoretical Physics at the Institute for Nuclear
Research RAS, where the results of the present study were
presented for the first time on 02.12.2013. The author also thanks
V.P. Gusynin, P.M. Lavrov, O. Lechtenfeld, P.Yu. Moshin and
K.V. Stepanyantz for useful discussions, as well as to I.V. Tyutin
for comments on the unitarity problem. I am grateful to the authors
of \cite{Sorellas} for their critical assessment of earlier papers,
leading to a better understanding of the problem under consideration.
I thank A.D. Pereira Jr. for discussions and comments on the Gribov
horizon in the Coulomb and maximal Abelian gauges. The study
was supported by the RFBR grant under Project No. 12-02-00121,
and by the grant of Leading Scientific Schools of
the Russian Federation under Project No. 88.2014.2.

%\newpage

\appendix

\section*{Appendix}
\section{On Solution of  Equation  (\ref{eqlambdapsi})}

\label{AppA} \renewcommand{\theequation}{\Alph{section}.\arabic{equation}} \setcounter{equation}{0}

In this Appendix, we present arguments for the existence of a
solution for Eq. (\ref{eqlambdapsi}) with respect to an unknown
field-dependent odd functional, $ \Lambda\left(\Phi,\Phi ^{\ast
}\right)$, in the form  (\ref{solcompeq}). In doing so, we follow
a strategy partially based on some previously known facts. First,
any gauge theory can be equivalently transformed to a gauge theory
in the standard basis \cite{blt}, with the generators and proper
zero eigenvectors having the representation \beq \label{standform}
\big\{R^i_{\alpha_0},Z^{\alpha_{0}}_{\alpha_{1}},\ldots,
Z^{\alpha_{L-2}}_{\alpha_{L-1}},
Z^{\alpha_{L-1}}_{\alpha_{L}}\big\} \to
\left\{\big(R^i_{\alpha},0\big), \left(\begin{array}{c|c}
                                                       0 & \delta^{\bar{\alpha}_0}_{B_{1}} \\
                                                       \hline
                                                       0 & 0
                                                     \end{array}\right)
,\ldots, \left(\begin{array}{c|c} 0 & \delta^{\bar{\alpha}_{L-2}}_{B_{L-1}} \\
                                                       \hline
                                                       0 & 0
                                                     \end{array}\right)
                                                     ,\big(0,\delta^{\bar{\alpha}_{L-1}}_{B_{L}}\big) \right\}
\eeq for the division of indices $\alpha _{s}$, $s=0,...,L$ being
related with the rank conditions (\ref{redth}), (\ref{redth1}) as
$\alpha _{0}=(\alpha ,B_{0})$, $\alpha
_{s}=\left(\bar{\alpha}_{s},B_{s+1}\right)$, for $s=1,\ldots ,L-1$
and $\alpha _{L}=B_{L}=m_L$. Note that the definition
(\ref{redth}) of an $L$-stage reducible gauge theory in the
standard basis (\ref{standform}) looks simple,
$Z^{\alpha_{s-1}}_{\alpha_{s}}Z^{\alpha_s}_{\alpha_{s+1}}= 0$, for
vanishing $K^{i\alpha_{s-1}}_{\alpha_{s+1}}$, for $s=0,...,L-1$.
Second, a transition to the standard basis from the initial gauge
theory can be realized as a non-degenerate (generally, non-local)
change of variables, $\Phi ^{A}\rightarrow \Phi ^{\prime A}(\Phi
)$, in $\mathcal{M}$, such that
\begin{eqnarray}
Z_{\Psi }(0,\Phi ^{\ast })& = & \int \!D\Phi \ \exp \Big\{%
\frac{\text{i}}{\hbar }S_{\Psi }\Big\}=\int \!D\Phi ^{\prime }\ \exp \Big\{%
\frac{\text{i}}{\hbar }\bar{S}_{\Psi }(\Phi ^{\prime })\Big\},\notag \\
&&\mathrm{with}\,\,\,\bar{S}_{\Psi }(\Phi ^{\prime })\ =\ S_{\Psi }(\Phi (\Phi
^{\prime }))-i\hbar \,\mathrm{Str}\,\ln\left\|\frac{\delta \Phi ^{A}}{\delta \Phi
^{\prime B}}\right\|.  \label{chvar1}
\end{eqnarray}%
We then use the fact that any gauge theory with an open algebra of
generators  $R^i_{\alpha}$ (being already in standard basis) can
be equivalently transformed to a theory with a closed algebra \cite%
{Voronovtyutin}, so that in the new basis of the generators of
gauge transformations, $R_{\alpha}^{\prime i}$ [obtained by means
of additive extension of $R^i_{\alpha}$ by trivial gauge
generators,
$R_{\alpha}^{\prime i}(A^{\prime}) = R_{\alpha}^{ i}(A^{\prime})+\mathcal{S}_{0,j}(A^{\prime}) M^{ij}_{\alpha}(A^{\prime})$],
the Lie-type structure functions $%
F_{\alpha\beta}^{\prime {\gamma}}(A^{\prime })$ in relations such as (%
\ref{commgtr}),
\begin{equation}
R_{\alpha}^{\prime i},_{j}(A^{\prime })R_{\beta}^{\prime j}(A^{\prime
})-R_{\beta}^{\prime i},_{j}(A^{ \prime })R_{\alpha}^{\prime
j}(A^{ \prime })=-R_{\gamma}^{\prime i}(A^{ \prime
})F_{\alpha\beta}^{\prime {\gamma}}(A^{\prime }),\ \left[\mathrm{ where} \  R_{\alpha}^{\prime i},_{j} \equiv \frac{\delta}{\delta A^{\prime j} } R_{\alpha}^{\prime i}\right] \label{commgtrn}
\end{equation}%
are the only ones to survive. A transition to the gauge theory subject to
relations (\ref{commgtrn}) may also be effectively realized as a non-degenerate change of
variables, $\Phi ^{\prime A}\rightarrow \Phi ^{\prime\prime A}(\Phi )$  in $\mathcal{M}$:%
\begin{eqnarray}
Z_{\Psi }(0,\Phi ^{\ast })& = & \int \!D\Phi ^{\prime }\
\exp \Big\{\frac{\text{i}}{\hbar }\bar{S}_{\Psi }(\Phi ^{\prime })\Big\}%
=\int \!D\Phi ^{\prime \prime }\ \exp \Big\{\frac{\text{i}}{\hbar }\hat{S}%
_{\Psi }(\Phi ^{\prime \prime })\Big\}, \notag \\
&&\mathrm{with}\ \,\,\hat{S}_{\Psi }(\Phi ^{\prime \prime })\ =\ \bar{S}_{\Psi
}(\Phi ^{\prime }(\Phi ^{\prime \prime }))-i\hbar \, \mathrm{Str}\,\ln \left\Vert\frac{%
\delta \Phi ^{\prime A}}{\delta \Phi ^{\prime \prime B}}\right\Vert.  \label{chvar2}
\end{eqnarray}%
Notice that the transformations $\Phi \rightarrow \Phi ^{\prime
}$, $\Phi ^{\prime }\rightarrow \Phi ^{\prime\prime }$ have a more
general form than the gauged BRST transformations (\ref{BRSTg})
and can be equivalently realized by a set of operations
(\ref{qaBV}) with definite respective functionals, $X_i(\Phi ,\Phi
^{\ast })$, for $i=1,2$ which convert a solution of the master
equation (\ref{MEBV}) into another solution $\hat{S}_{\Psi }$,
\begin{equation}
\hat{S}_{\Psi }=\frac{\hbar }{i}\,\mbox{ln}\,\left[ \exp \left\{ -[\Delta
,X_2]\right\}\cdot \exp \left\{ -[\Delta
,X_1]\right\} \exp \left\{ \frac{\text{i}}{\hbar }{{S}_{\Psi }}\right\} %
\right] \label{x1x2}.
\end{equation}

Since the transformed action $\hat{S}_{\Psi }$ (\ref{chvar2}) has
a form being linear in the antifields, $\hat{S}_{\Psi }(\Phi
^{\prime \prime },\Phi ^{\ast })=\Phi _{A}^{\ast }\hat{S}_{\Psi
}^{A}(\Phi ^{\prime \prime })$, we now obtain the relations
(derivatives with respect to the fields in $\hat{S}_{\Psi
}^{A},_{B}$ and $\hat{S}_{\Psi },_{B}$ are understood as taken for
$\Phi ^{\prime \prime B}$, and we omit the Jacobi  matrices of the
above changes of variables for the sake of simplicity)
%%%%%%%%%
\beq\label{conseqhatS} \big(\hat{S}^{AB}_\Psi\ =\ 0, \quad \Delta
\hat{S}^{A}_\Psi\ =\ 0\big) \Longrightarrow \hat{S}^{A}_\Psi,_B
\hat{S}^{B}_\Psi\ =\ 0, \eeq
%%%%%%%%%%%
which, first of all, imply the nilpotency of the Slavnov variation, $%
\hat{s}_{e}^{2}=0$, in the new basis of the gauge algebra and,
second, allow one to present the equation (\ref{eqlambdapsi}) for
a gauge theory with a closed
algebra as an equation for the parameter $\hat{\Lambda}$,\footnote{%
For simplicity, we use notation for the gauge fermion $\Psi $ and
its variation $\overline{\Delta}\Psi$ in the case of a theory with
a closed algebra which is the same as the notation used for a
theory with an open algebra and the action ${S}_{\Psi }$.}
%%%%%%%%
 \beq
\label{eqlambdapsiclalg}
 i\hbar\left\{\ln \big(1+\hat{s} \hat{\Lambda}\big)\right\}\  =\ \hat{s} \bigl(\overline{\Delta}\Psi(\Phi'')\big),
\eeq where account has been taken of the fact that the generator
$\hat{s}$ coincides with $\hat{s}_{e}$, being, however, expressed
in terms of the action $\hat{S}_{\Psi }$ and fields $\Phi ^{\prime
\prime A}$.

Using the functional equation (\ref{eqlambdapsiclalg}), we can express the
variation $\overline{\Delta }\Psi (\Phi ^{\prime \prime })$ with accuracy up
to BRST exact terms, $\hat{s}R(\Phi ^{\prime \prime })$,
\begin{equation}\label{hatdpsi}
\overline{\Delta }\Psi (\Phi ^{\prime \prime })\ =\ i\hbar \hat{\Lambda}%
(\Phi ^{\prime \prime })(\hat{s}\hat{\Lambda})^{-1}\left\{ \,\mbox{ln}\,\big(1+\hat{s}%
\hat{\Lambda}\big)\right\} ,
\end{equation}%
which is identical with the variations for finite field-dependent BRST transformations
in the Yang--Mills theory \cite{ll1}, now proved to be valid for a theory
with a closed algebra. A solution of Eq. (\ref{eqlambdapsiclalg})
with respect to an unknown $\hat{\Lambda}(\Phi ^{\prime \prime })$ reads as follows:%
\begin{eqnarray}
&&\hat{s}\hat{\Lambda}(\Phi ^{\prime \prime })\ =\ \exp \left\{ -\frac{i}{\hbar }\hat{s}%
\big(\overline{\Delta }\Psi (\Phi ^{\prime \prime })\big)\right\}
-1 \ \Longrightarrow  \notag  \label{aollambda} \\
&&\hat{\Lambda}(\Phi ^{\prime \prime })\ =\ \overline{\Delta }\Psi (\Phi
^{\prime \prime })\big(\hat{s}\overline{\Delta }\Psi (\Phi ^{\prime \prime })\big)%
^{-1}\Big[\exp \left\{ -\frac{i}{\hbar }\hat{s}\big(\overline{\Delta }\Psi (\Phi
^{\prime \prime })\big)\right\} -1\Big].
\end{eqnarray}

Finally, in order to obtain a solution of the initial equation (\ref%
{eqlambdapsi}) which equivalently may be rewritten as\footnote{In (\ref{eqlambdapsifin}) the action of group-like element  $g(\Lambda(\Phi,\Phi^*))=\big(1+\overleftarrow{s}_{e}\Lambda \big)$,  being trivial for nilpotent $s_{e}$, measures difference of (\ref%
{eqlambdapsi}) with the equation (\ref{eqlambdapsiclalg}) for the gauge theory with closed algebra}
\begin{eqnarray}
&&  \big(1+s_{e}\Lambda \big)^{-1} \big(1+\overleftarrow{s}_{e}\Lambda \big)  = \exp\left[\frac{i}{\hbar}\Big(\exp\Big\{ -[\Delta,\,\overline{\Delta} \Psi] \Big\}-1\Big)S_\Psi\right] \label{eqlambdapsifin}
\end{eqnarray}%
  we have to make the inverse transformations
$\Phi ^{\prime\prime } \rightarrow \Phi ^{\prime } \rightarrow \Phi $ for $\hat{\Lambda}%
(\Phi ^{\prime \prime })$ with respect to those used for the
transition to the standard basis (\ref{standform}) and then to the
gauge theory with a closed algebra (\ref{commgtrn}), described in Eqs. (\ref{chvar1}), (\ref%
{chvar2}), and therefore a solution, ${\Lambda }(\Phi ,\Phi ^{\ast
})$, of Eq. (\ref{eqlambdapsi}) does exist and is expressed by the variation $%
\overline{\Delta }\Psi (\Phi )$ in the form (\ref{solcompeq}).

\end{document}